\documentclass[12pt,a4paper]{article}
\usepackage{amsmath,caption,amsfonts,setspace,mathrsfs,mathtools,bbm}
\usepackage[top=1.25in, bottom=1.25in, left=1.2in, right=1.2in]{geometry}

\usepackage[round]{natbib}
\usepackage{color,soul}
\renewcommand{\baselinestretch}{1.4} 
\DeclareCaptionStyle{italic}[justification=centering]{labelfont={bf},textfont={it},labelsep=colon}
\usepackage{graphicx,psfrag,epsf,textcomp}
\usepackage{enumerate, dsfont}
\usepackage{url,xcolor} 
\usepackage{booktabs, subfig, bm, paralist,mathpazo,tikz,todonotes,longtable,microtype}
\usepackage{mathrsfs}
\usepackage{calligra,amssymb}
\usepackage[pdftex,colorlinks=true]{hyperref}
\definecolor{darkblue}{rgb}{0,0,.6}
\hypersetup{citecolor=darkblue,linkcolor=darkblue,urlcolor=darkblue}

\newcommand{\blind}{1}

\graphicspath{{plots/}}
\DeclareMathOperator*{\argmin}{\arg\!\min}
\newsavebox\CBox

\definecolor{a0}{rgb}{0.0, 0.5, 0.0}
\definecolor{bistre}{rgb}{0.24, 0.17, 0.12}
\definecolor{amethyst}{rgb}{0.6, 0.4, 0.8}
\definecolor{blue-violet}{rgb}{0.54, 0.17, 0.89}
\definecolor{Rcolor}{RGB}{150,160,190}
\definecolor{blush}{rgb}{0.87, 0.36, 0.51}
\definecolor{brightturquoise}{rgb}{0.03, 0.91, 0.87}
\definecolor{burntorange}{rgb}{0.8, 0.33, 0.0}
\usepackage{orcidlink}

\date{}

\begin{document}

\def\spacingset#1{\renewcommand{\baselinestretch}{#1}\small\normalsize} \spacingset{1}

\if0\blind
{
  \title{\bf Is the group structure important in grouped functional time series?}
  \maketitle
} \fi

\if1\blind
{
    \title{\bf Is the group structure important in grouped functional time series?}
    \author{Yang Yang \orcidlink{0000-0002-8323-1490} \\ 
    Department of Econometrics and Business Statistics \\
    Monash University \\
    \\
    Han Lin Shang \orcidlink{0000-0003-1769-6430} \footnote{Postal address: Department of Actuarial Studies and Business Analytics, Level 7, 4 Eastern Road, Macquarie University, Sydney, NSW 2109, Australia; Telephone: +61(2) 9850 4689; Email: hanlin.shang@mq.edu.au} \\
     Department of Actuarial Studies and Business Analytics \\
     Macquarie University}
  \maketitle
} \fi

\begin{abstract}
We study the importance of group structure in grouped functional time series. Due to the non-uniqueness of group structure, we investigate different disaggregation structures in grouped functional time series. We address a practical question on whether or not the group structure can affect forecast accuracy. Using a dynamic multivariate functional time series method, we consider joint modeling and forecasting multiple series. Illustrated by Japanese sub-national age-specific mortality rates from 1975 to 2016, we investigate one- to 15-step-ahead point and interval forecast accuracies for the two group structures.

\vspace{0.2cm}

\noindent \textit{Keywords}: dynamic principal component analysis; forecast reconciliation; Japanese sub-national age-specific mortality rates; long-run covariance function; multivariate functional principal component analysis

\end{abstract}

\newpage
\doublespacing

\section{Introduction}\label{sec:intro}

Because of the rapid advancements in medical technology and improvements in health services, many countries worldwide have been observing a continuous reduction of human mortality rates for decades. Developed OECD nations, such as Japan, consider the remarkable increase in population life expectancy as one of the greatest achievements of the last century \citep{OECD19}. The growing longevity risks in many countries put pressure on government policymakers and planners to develop sustainable pension, health, and aged care systems \citep[see, e.g.,][]{Coulmas07}. Traditional studies of human mortality focus on population data at the national level. In recent years, academics and government stakeholders have shown increasing interest in regional mortality improvements after realizing that sub-national forecasts of age-specific mortality rates help inform social and economic policies within local regions. Thus, any improvement in the forecast accuracy of mortality rates will be beneficial in determining the allocation of current and future resources at the national and sub-national levels. 

Numerous mortality modeling methods have been proposed since the first publication of the Gompertz law in 1825 \citep[see][for a comprehensive literature review on mortality modeling and forecasting]{Booth2008}. Only a few approaches could simultaneously forecast multiple regional mortality rates within a country \citep[see, e.g.,][]{JK11, CBD+11, DCB+11, RGH11, HH13, Shang16c, VHK+17, Gao2019, PRS+21}. The main hypothesis for multi-population mortality models is that mortality rate differences for any two populations having similar socio-economic status and close connections with each other do not diverge indefinitely over time. This paper applies a dynamic multivariate functional time series method to produce accurate sub-national age-specific mortality forecasts.

Applying functional time series forecasting methods to sub-national mortality disaggregated by attributes such as sex and geographical regions independently does not ensure coherence in the forecasts. The forecasts of all sub-populations will not add up to the forecasts obtained by applying the technique to the national data. Hence, in practice, it is crucial to consider forecast reconciliation approaches in forecasting sub-national mortality rates \citep[e.g., see][]{SCM42, SW09}. 

We apply a dynamic multivariate functional time series method to simultaneously model a group of sub-national mortality rates and make forecasts. In particular, we aim to answer a practical issue: does the group structure matter in terms of forecast accuracy for a grouped functional time series? We first arrange the national and sub-national Japanese male and female age-specific mortality rates from 1975 to 2016 into two group structures. Then, we apply dynamic univariate and multivariate functional time series methods and compare their forecasting performance on some holdout data for the two group structures considered.

The rest of this paper is structured as follows. In Section~\ref{sec:data}, we describe the motivating dataset, which is the Japanese national and sub-national age-specific mortality rates. In Section~\ref{sec:method}, we describe a dynamic multivariate functional time series method for producing point and interval forecasts. Reconciled forecasts produced by the dynamic univariate and multivariate functional time series forecasting methods are evaluated and compared in Section~\ref{sec:result}.  Conclusions are presented in Section~\ref{sec:conclusion}, along with some reflections on ways to extend the techniques presented in this study. 

\section{Japanese age-specific mortality rates} \label{sec:data}

\begin{figure}[!htbp]
\centering
\subfloat[Observed female mortality rates.]
{\includegraphics[width = 3.34in]{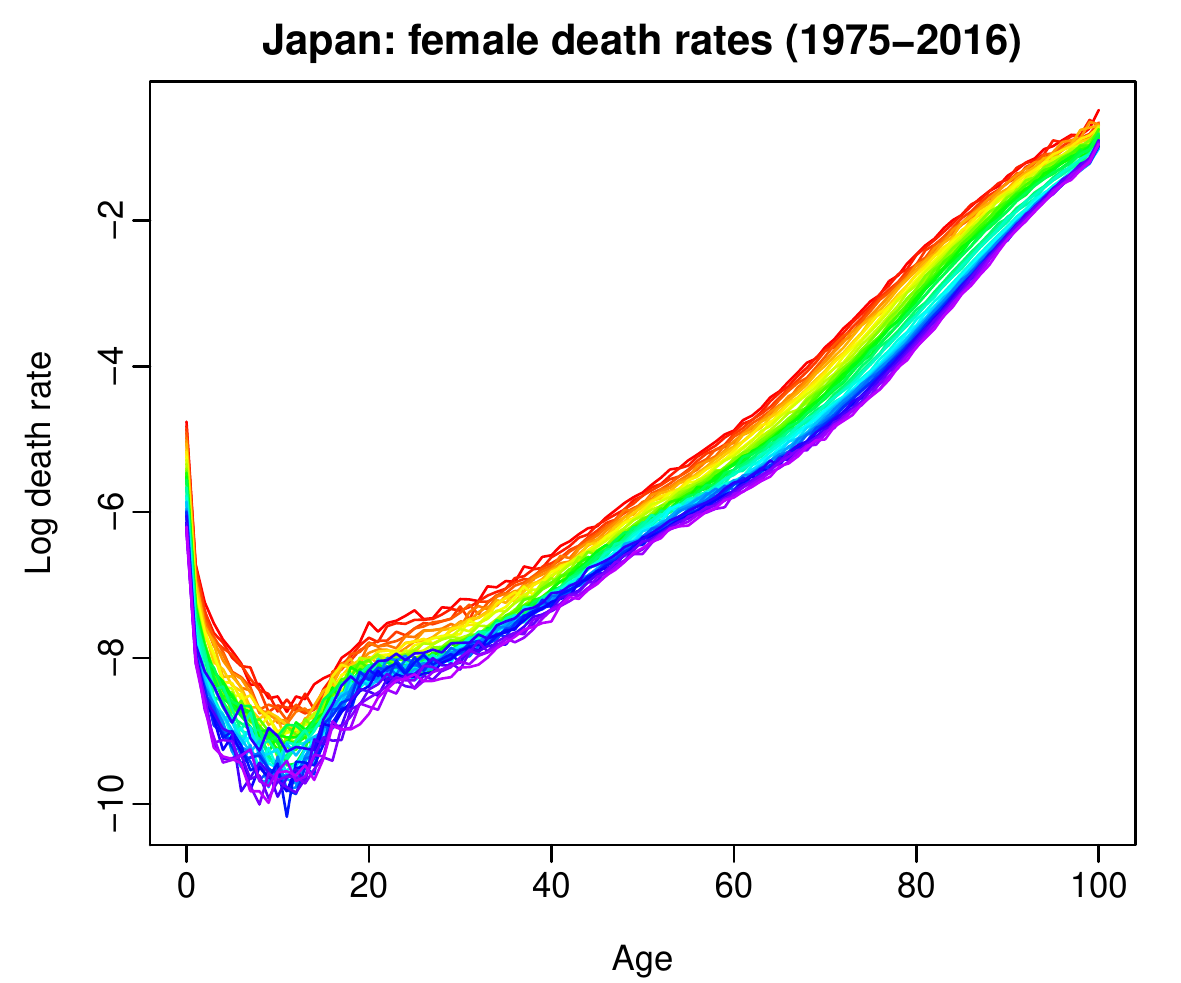}} \quad
\subfloat[Observed male mortality rates.]
{\includegraphics[width = 3.34in]{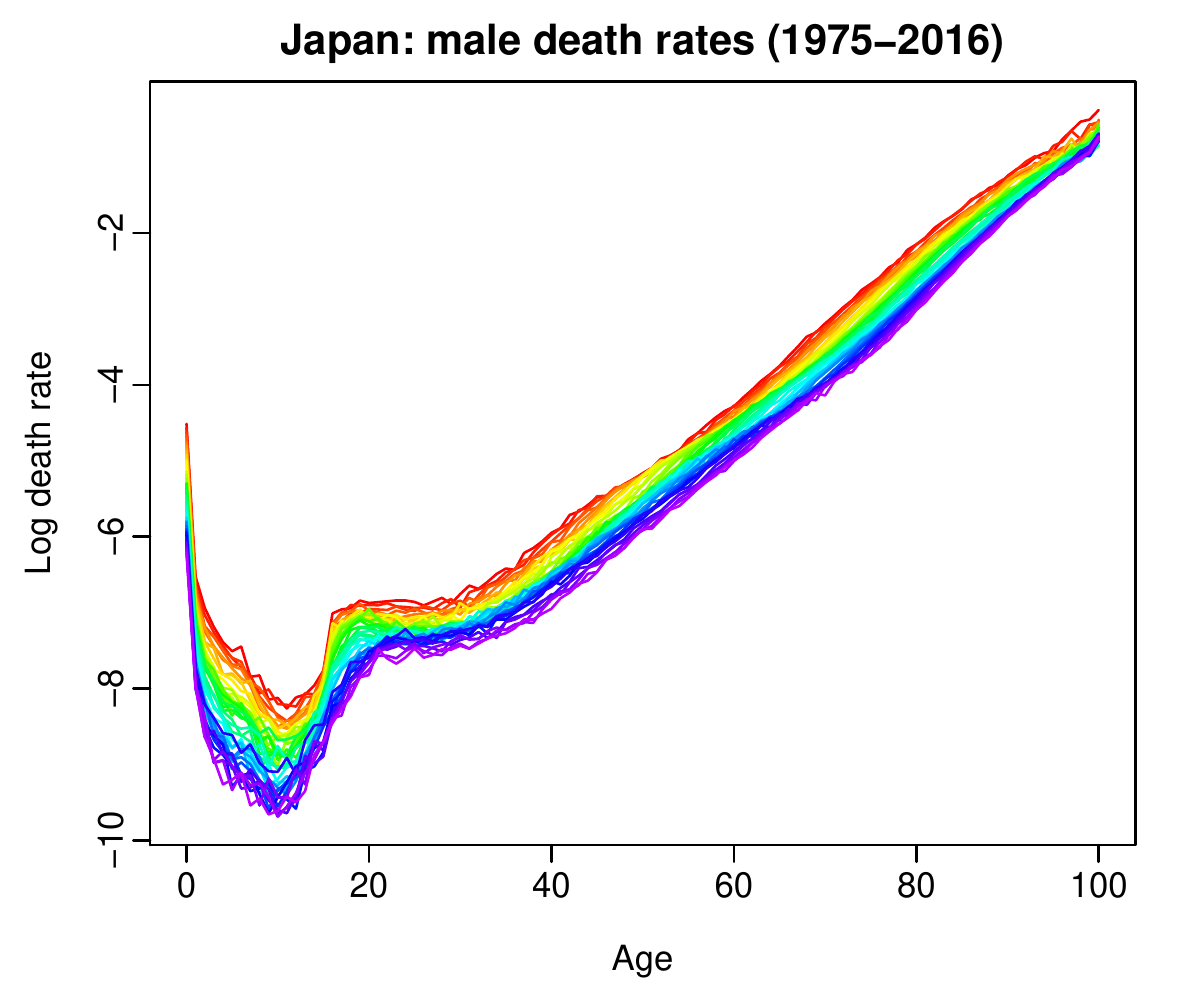}} \\
\caption{Functional time series graphical displays for the Japanese age-specific mortality rates from 1975 to 2016. Using a rainbow color palette, the mortality rates in the distant past years are shown in red, while the mortality rates in the most recent years are shown in purple.}\label{fig:1}
\end{figure}

The Japanese Mortality Database regularly recently publishes age-specific mortality rates over the period 1975--2016 at the national level and various subnational levels \citep{JHMD16}. We consider ages from 0 to 99 in single years of age and include the last age group, ``100+'', as an aggregation of ages at and beyond 100. Figure~\ref{fig:1} presents rainbow plots of the age-specific mortality rates of females and males in the studied period. Rainbow plots are designed to display the distant past in red curves and the more recent years in purple curves \citep{HS10}. Typical ``J shape'' mortality patterns of developed countries can be observed in both panels in Figure~\ref{fig:1}, which indicates a rapid decrease of mortality occurs for the newborns, with slower reduction continuing into the adolescence age. There is an ``accident hump'' in mortality for young adults around age 20, followed by a short ``plateau'' until the early 30s, and then a steady increase in middle-to-old age groups. Figure~\ref{fig:1} reveals an overall reduction in the trend of total mortality rates over the considered period 1975--2016. However, the rate of decline appears to vary across different age groups.

\subsection{Nonparametric smoothing}

We implement a penalized regression spline smoothing method with a monotonic constraint to smooth out excess noise at higher ages \citep[see also][]{HU07, Shang16c}. 

For a particular population $j$, let $y_t^j(x_i)$ be the log central mortality rates observed at the beginning of each year for year $t=1,2,\dots,n$, where $x_i$ represents the center of each age or age group for $i = 1, \dots, p$. The value of $j$ depends on the research question. For example, when modeling age-specific mortality rates of females and males in a particular prefecture, $j$ takes two values: male or female. We assume there is an underlying continuous and smooth function $f_t^j(x)$ that is observed at discrete data points with errors of the form
\begin{equation*}
y_t^j(x_i) = f_t^j(x_i) + \delta_t^j(x_i) \varepsilon_{t,i}^j,
\end{equation*}
where $\varepsilon_{t,i}^j$ is an independent and identically distributed standard normal random variable for each age in year $t$, and $\delta_t^j(x_i)$ measures the variability in mortality at each age in year $t$ for the $j^{\text{th}}$ population. Jointly, $\delta_t^j(x_i)\varepsilon_{t,i}^j$ represents the smoothing error.

Let $m_t^j(x_i) = \exp\{y_t^j(x_i)\}$ be the observed central mortality rates for age $x_i$ in year $t$ and define $E_t^j(x_i)$ to be the total exposure to risks of $j^{\text{th}}$ population of age $x_i$ at 1\textsuperscript{st} January of year $t$. 
Then $m_t^j(x_i)$ approximately follows a binomial distribution with estimated variance $\left(E_t^j(x_i)\right)^{-1}m_t^j(x_i) \left[1-m_t^j(x_i)\right] $. Since we are modeling the mortality rates at logarithm scale, we apply a Taylor approximation to $y_t^j(x) = \ln(m_t^j(x_i))$ to get
\begin{equation*}
\left(\widehat{\delta}_t^j\right)^2(x_i) := \text{Var}\left\{\ln [m_t^j(x_i)]\right\} \approx \frac{1-m_t^j(x_i)}{m_t^j(x_i)\times E_t^j(x_i)}.
\end{equation*}
Following \cite{HU07}, we then define weights $w_t^j(x_i) = 1/\left(\widehat{\delta}_t^j\right)^2(x_i)$ and use weighted penalized regression splines to obtain the smoothed functions $f_t^j(x)$ \citep[see also][]{Wood2003,CDE04}.

\begin{figure}[!htbp]
\centering
\subfloat[Smoothed female mortality rates.]
{\includegraphics[width = 3.34in]{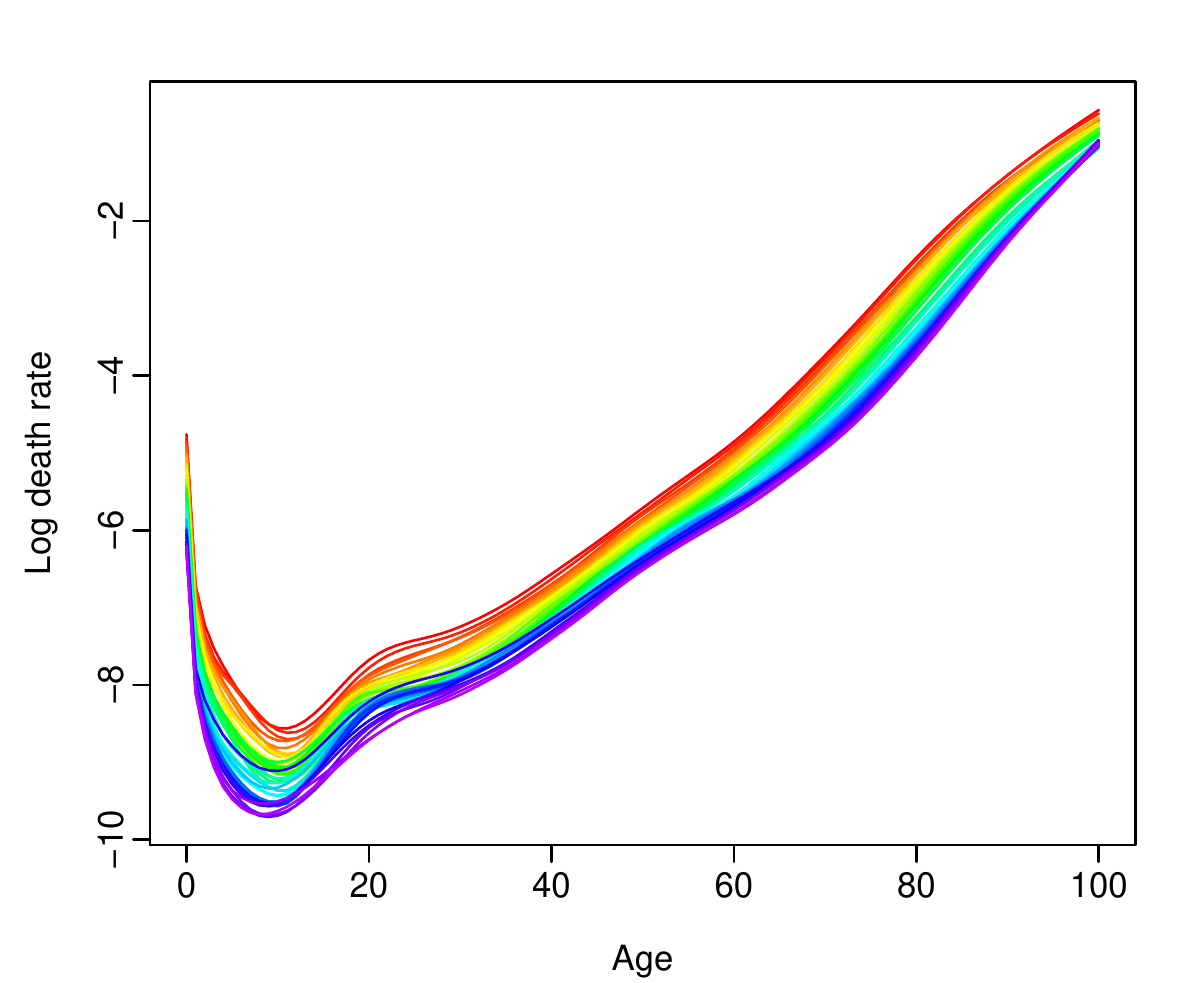}}\quad
\subfloat[Smoothed male mortality rates.]
{\includegraphics[width = 3.34in]{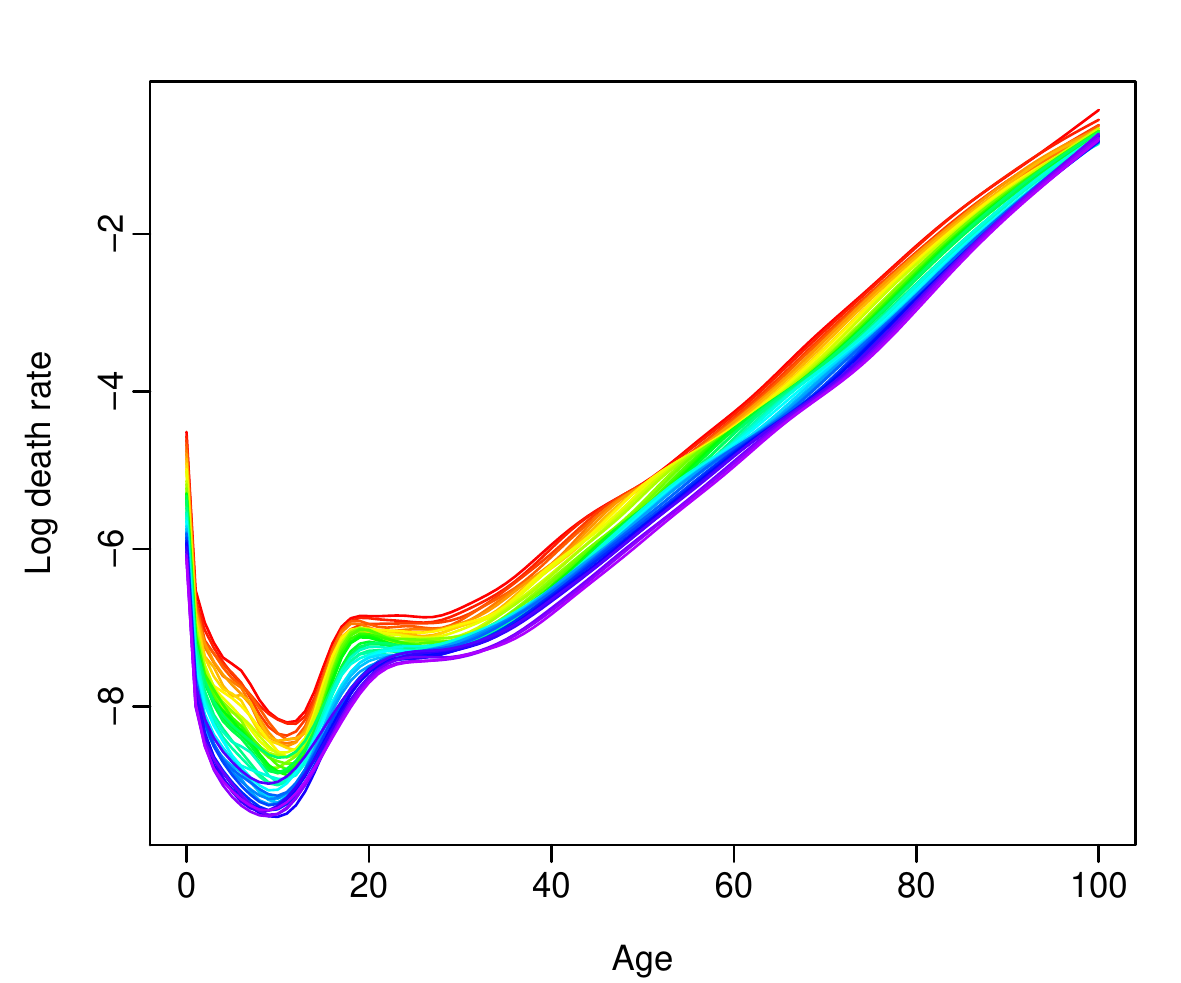}} 
\caption{Functional time series graphical displays for the smoothed series, where a penalized regression spline smoothing with monotonic constraint is implemented.}\label{fig:2}
\end{figure}

As a byproduct of the smoothing, we can potentially fill in any missing observations. In Figure~\ref{fig:2}, we present the smoothed log mortality curves.

\begin{figure}[!htb]
\centering
{\includegraphics[width = 3.1in]{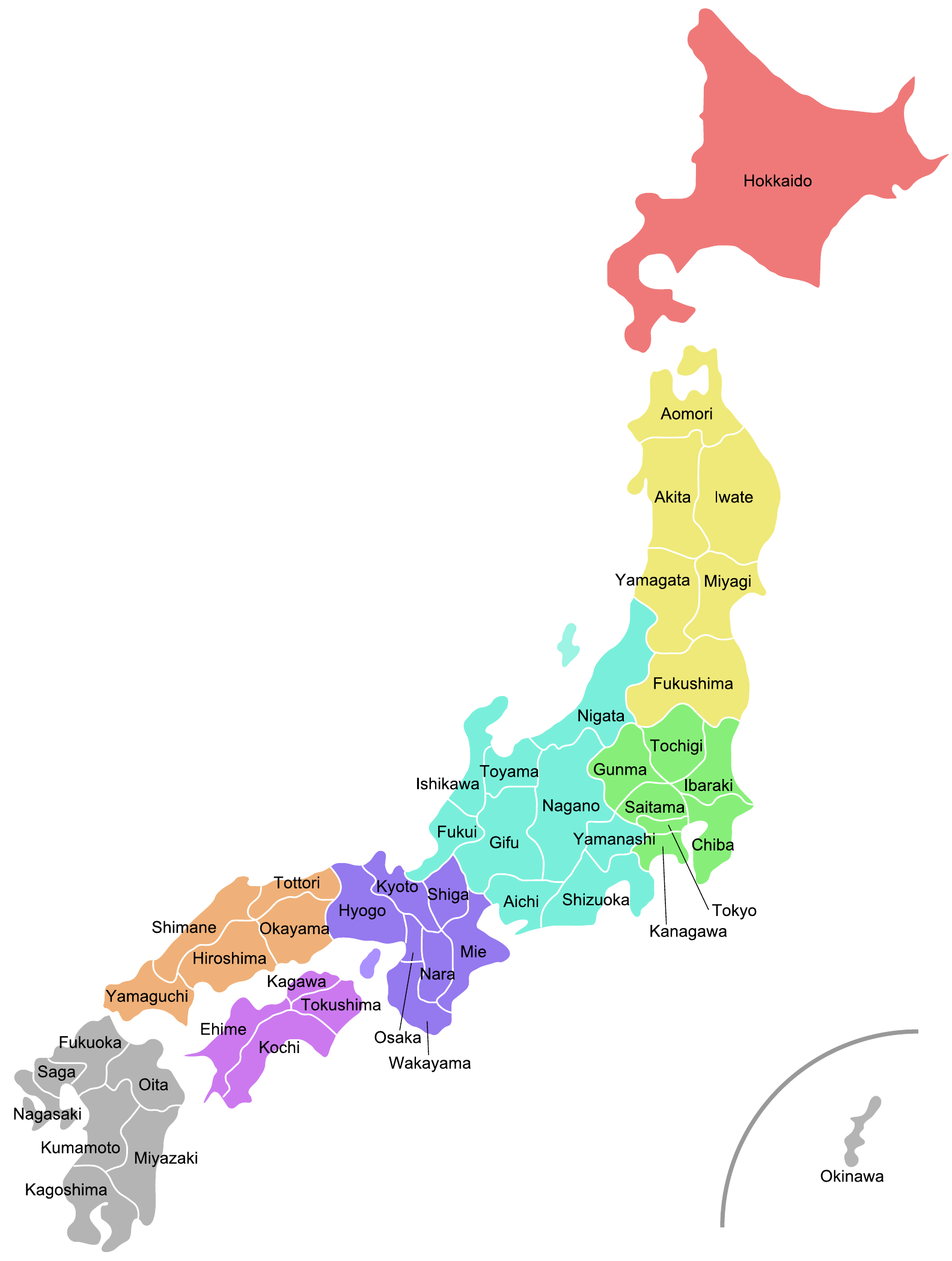}} 
\caption{Eight regions (represented in different colors) and 47 prefectures of Japan. In the remaining sections, we mainly use notations $R_1,\cdots, R_8$, and $P_1,\cdots, P_{47}$ numbered geographically from north to south when referring to regions and prefectures, respectively.} \label{fig:3}
\end{figure}

To better understand the mortality patterns of Japan, we consider mortality time series featuring various geographical factors (i.e., country, region, and prefecture) and sex factors (i.e., female and male) in this study. As highlighted by different colors in Figure~\ref{fig:3}, Japan can be divided into eight geographical regions, namely, Hokkaido (red), Tohoku (yellow), Kanto (green), Chubu (cyan), Kansai (purple), Chugoku (orange), Shikoku (magenta), and Kyushu (grey). Regions vary not only in geography locations, sizes, climates, and terrains but also in histories, traditions, and cultures \citep{JG}. Apart from the eight regions, there are 47 prefectures in Japan, forming the country's base level of administrative division. Thus, disaggregate the national and sub-national age-specific mortality rates by geographical location and sex leading to in total 168 series (i.e., $47\times 2 = 94$ Sex $\times$ Prefecture series, 16 Sex $\times$ Region series, 2 Sex series at the national level, and further $1+8+47 = 56$ total series at Japan, Region and Prefecture levels, respectively). Because of the non-uniqueness of the group structure, the middle-level series can be disaggregated differently. This, in turn, affects the form of a group structure. 

\begin{figure}[!htb]
	\centering
	\subfloat[Geographical hierarchy]
	{\includegraphics[width = 3in]{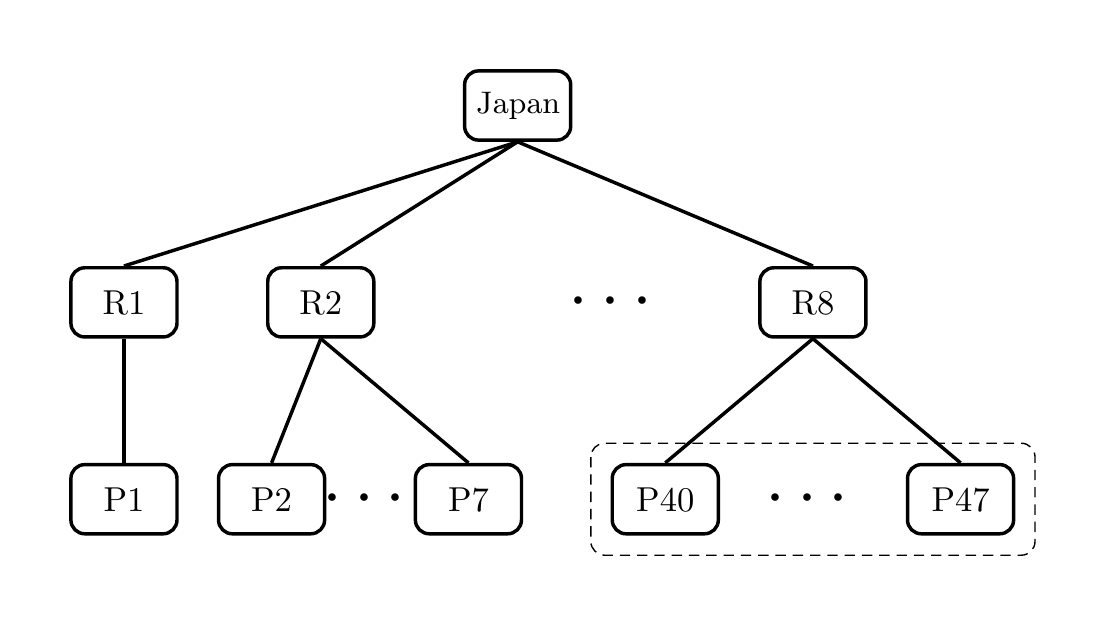}\label{fig:4a}} \qquad
	\subfloat[Hierarchy 1]
	{\includegraphics[width = 3in]{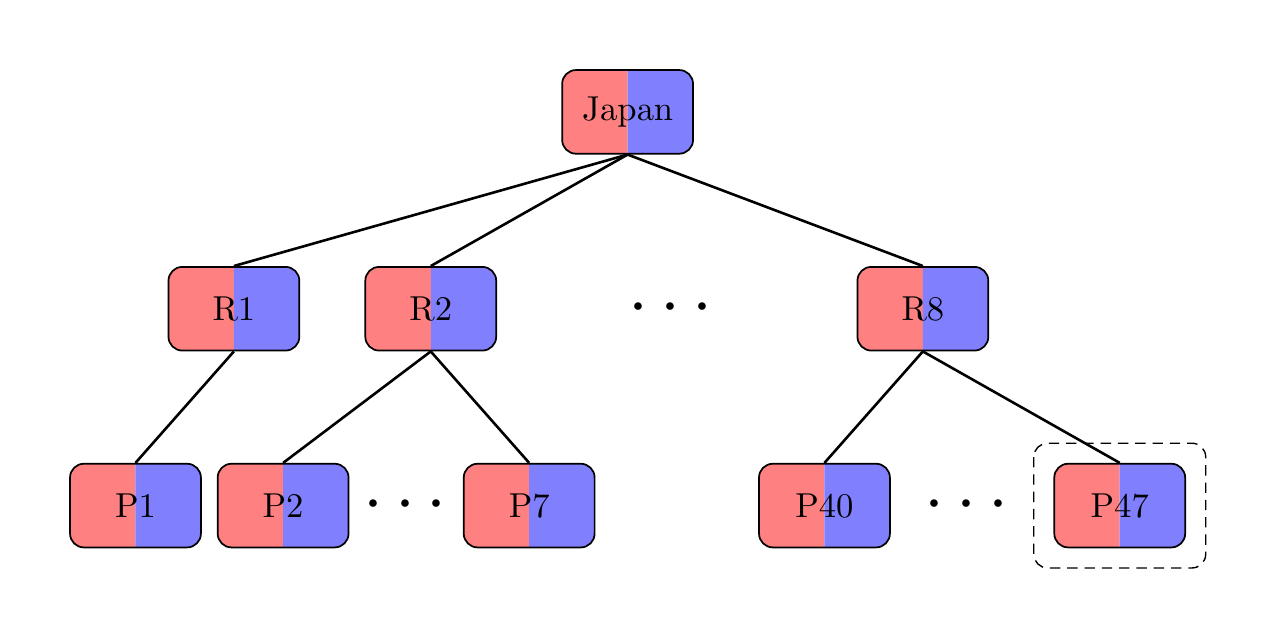}\label{fig:4b}} \\
	\subfloat[Hierarchy 2]
	{\includegraphics[width = 6in]{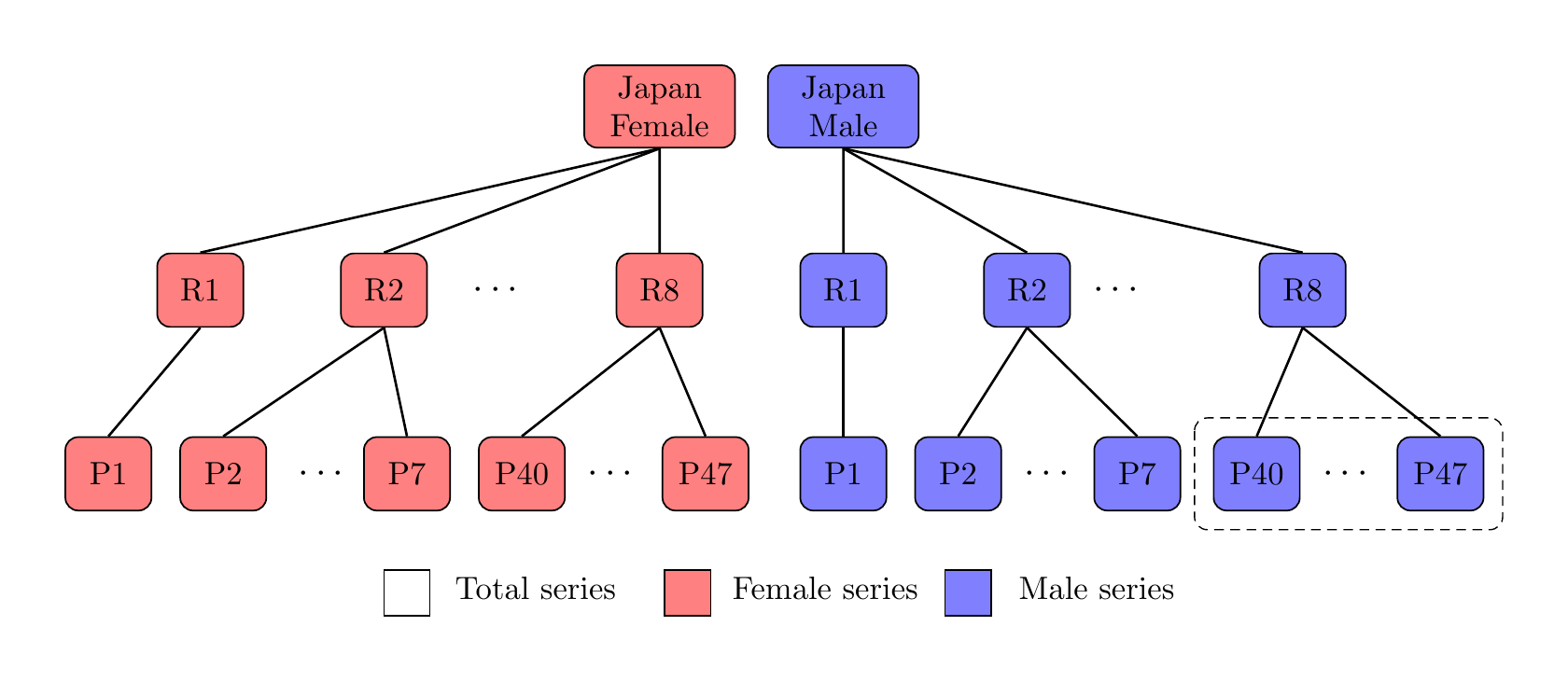}\label{fig:4c}} 
	\caption{Hierarchy tree diagrams for the Japanese mortality rates. Dashed boxes show examples of jointly modeling multiple series under various hierarchy structures.}\label{fig:4}
\end{figure}

We split the Japanese data into geographical structures coupled with a sex grouping variable by considering the difference in mortality between females and males. A three-level geographical structure is used to disaggregate the national and sub-national total mortality rates, as shown in Figure~\ref{fig:4}\subref{fig:4a}. We apply a dynamic multivariate functional time series method to eight region-level series and 47 prefecture-level series, respectively, to obtain joint forecasts at each level. Specifically, for each region, we consider all prefecture-level total series that belong to this region; all eight region-level total series are jointly modeled for the national total. These forecasts of multiple series at various levels are then used to perform forecast reconciliation according to the specified group structure. 

When forecasting mortality rates of females and males, two disaggregation structures focusing on either geographical factors or sex are used. The first grouping structure (Hierarchy~1) in Figure~\ref{fig:4}\subref{fig:4b} initially disaggregates the top-level series by geographical location. It then further splits sub-national totals into sex-specific series at the region and prefecture levels. Jointly modeling the two sex-specific series at each node of the hierarchy incorporates the close connection of females and males living in each region or prefecture. The second grouping structure (Hierarchy~2) in Figure~\ref{fig:4}\subref{fig:4c} first splits the total series by sex and then further disaggregates the totals for females and males according to geographical factors. Implementing a dynamic multivariate functional time series method under this structure means jointly modeling the series of one particular sex (female or male at a time) either for each region or for the national sex-specific totals. This sub-national mortality forecasting considers the covariance between spatially closely correlated sub-populations and respects the well-known fact that human females and males have different mortality patterns. 

\section{Methodology}\label{sec:method}

The mortality patterns of sub-populations in Japan are expected to be similar because of their closely related biological backgrounds and similar lifestyles. Still, they are not expected to be identical nationwide across all age groups. Traditional age-specific mortality forecasting methods tend to result in divergent forecasts for sub-populations when independence is assumed. It is desirable to model sub-national mortality rates simultaneously while considering the potential heterogeneity between these. In this paper, we apply a dynamic multivariate functional time series method and demonstrate its advantage over its univariate counterpart using the Japanese sub-national data in Section~\ref{sec:result}.

\subsection{Dynamic multivariate functional time series method} \label{sec:3.1}

The dynamic univariate functional time series method considers temporal dependence of data by estimating its long-run covariance function. Let $\{ f^{(l)}(x) \}_{l = 1, \dots, \omega }$ be a set of random functions (in total we have $\omega \in \mathbb{Z}^{+}$ functions with $\omega\geq 2$) that comprise each observation in the considered data set. Each $f^{(l)}(x)$  is defined in $\mathcal{L}^2(\mathcal{I})$, a Hilbert space of square-integrable functions on an interval $\mathcal{I} = [a,b]$, $a<b$. Write $\bm{f}(x) = \left[f^{(1)}(x),\dots,f^{(\omega)}(x)\right]^{\top}$ as a vector in a Hilbert space, consisting of $\omega$ number of functions. These multivariate functions may be defined on the same domain $\mathcal{I}$ \citep[see, e.g.,][]{JP14, CCY14} or may be defined at different domains $\mathcal{I}_1,\dots,\mathcal{I}_{\omega}$ \citep[see, e.g.,][]{HG16}. Given our application has the same domain, we consider the former situation.

We define the mean function
\begin{equation*}
\bm{\mu}(x) := \left[\mu^{(1)}(x), \dots, \mu^{(\omega)}(x) \right] = \left\{\text{E}\left[f^{(1)}(x)\right], \dots, \text{E}\left[f^{(\omega)}(x)\right]\right\}.
\end{equation*}
For $x, s\in \mathcal{I}$, auto-covariance function at lag $\ell$ is defined as 
\begin{equation*}
\gamma_{\ell} (x,s) := \text{Cov} \left[\bm{f}_0(x) , \bm{f}_{\ell}(s) \right] =  \text{E}\left[\left(\bm{f}_0(x) - \bm{\mu}(x) \right), \left(\bm{f}_{\ell}(s) - \bm{\mu}(x) \right)\right],
\end{equation*}
with elements 
\begin{equation*}
  \gamma^{(lj)}_{\ell}(s,t) := \text{E}\left[\left(f_0^{(l)}(x) - \text{E}\left[f^{(l)}(x)\right] \right) \left(f_{\ell}^{(j)}(s) - \text{E}\left[f^{(j)}(s)\right] \right)\right].
\end{equation*}
The long-run covariance is defined as
\begin{equation}
  C(x,s) := \sum_{\ell = -\infty}^{\infty} \gamma_{\ell} (x,s), \label{eq_1}
\end{equation}
where $f^{(j)}(s)$ and $f^{(l)}(x)\in \mathcal{L}^2(\mathcal{I})$.  The long-run covariance function $C(x, s)$ $C(x, s)$ consists of the variance as well as the temporal cross-covariance at various positive and negative lags of a functional time series, and is a wall-defined element of $\mathcal{L}^2(\mathcal{I})$ under mild dependence and moment conditions. Through right integration, $C(x, s)$ defines a Hilbert-Schmidt integral operator on $\mathcal{L}^2(\mathcal{I})$ given by 
\begin{equation*}
(\mathcal{K}(f))^{(j)}(s) = \sum^{\omega}_{l=1}\int_{\mathcal{I}}C_{lj}(x,s)f^{(l)}(x)dx,
\end{equation*}
where $\mathcal{K}$ denotes a kernel operator $\mathcal{L}^2(\mathcal{I}) \rightarrow \mathcal{L}^2(\mathcal{I})$ and $C_{lj}(x,s)$ is the $(l,j)\textsuperscript{th}$ element of $C(x,s)$.

Via Mercer's lemma, there exists an orthonormal sequence $(\phi_k)$ of continuous functions in $\mathcal{L}^2(\mathcal{I})$ and a non-increasing sequence $\lambda_k$ of positive number, such that
\begin{equation*}
C_{lj}(x,s) = \sum^{\infty}_{k=1}\lambda_k \phi_k^{(l)}(x)\phi_k^{(j)}(s).
\end{equation*}
By the separability of Hilbert spaces, the Karhunen-Lo\`{e}ve expansion of a stochastic process $\bm{f}^{(l)}(x) = \big\{f_1^{(l)}(x),\dots,f_n^{(l)}(x)\big\}$ can be expressed as
\begin{align}
f_t^{(l)}(x) &= \mu^{(l)}(x) + \sum^{\infty}_{k=1}\beta^{(l)}_{t,k}\phi_k^{(l)}(x) \notag\\
&= \mu^{(l)}(x) + \sum_{k=1}^{K}\beta^{(l)}_{t,k}\phi^{(l)}_k(x) + e^{(l)}_t(x), \label{eq_2}
\end{align}
where $\mu^{(l)}(x)$ is the mean function for the $l^{\text{th}}$ sub-population; $\Big\{\phi_1^{(l)}(x),\dots,\phi_{K}^{(l)}(x)\Big\}$ is a set of the first $K$ functional principal components; $\bm{\beta}_1^{(l)} = \left(\beta_{1,1}^{(l)},\dots,\beta_{n,1}^{(l)}\right)^{\top}$ and $\Big\{\bm{\beta}_1^{(l)},\dots,\bm{\beta}^{(l)}_{K}\Big\}$ denotes a set of principal component scores for the $l^{\text{th}}$ sub-population in~\eqref{eq_1}; $e_t^{(l)}(x)$ denotes the model truncation error function with mean zero and finite variance for the $l^{\text{th}}$ sub-population;  and $K<n$ is the number of retained principal components. Expansion~\eqref{eq_2} facilitates dimension reduction as the first $K$ terms often provide a good approximation to the infinite sums, and thus the information contained in $\bm{f}^{(l)}(x)$ can be adequately summarized by the $K$-dimensional vector $\left(\bm{\beta}^{(l)}_1,\dots,\bm{\beta}^{(l)}_{K}\right)$. Truncating at the first $K^{\text{th}}$ functional principal components gives the approximation in its matrix formulation as
\begin{equation*}
\bm{f}_t(x) = \bm{\Phi}\bm{\beta}_t^{\top},
\end{equation*}
where $\bm{f}_t(x) = \left[f_t^{(1)}(x),\dots,f_t^{(\omega)}(x)\right]$, $\bm{\beta}_t = \left(\beta_{t,1}^{(1)}, \dots, \beta_{t,K}^{(1)}, \beta_{t,1}^{(2)},\dots,\beta_{t,K}^{(2)},\dots, \beta_{t,1}^{(\omega)},\dots,\beta_{t,K}^{(\omega)}\right)$ being the vector of the basis expansion coefficients, and 
\begin{equation*}
\bm{\Phi}(x) = \left( \begin{array}{ccccccccc}
\phi_1^{(1)}(x) & \cdots & \phi_{K}^{(1)}(x) & 0 & \cdots & 0 & 0 & \cdots & 0 \\
0 & \cdots & 0 & \phi_1^{(2)}(x) & \cdots & \phi_{K}^{(2)}(x) & 0 & \cdots & 0 \\
\vdots & \vdots & \vdots & \vdots & \vdots & \vdots & \vdots & \vdots & \vdots \\
 0 & \cdots & 0 & 0 & \cdots & 0 & \phi_1^{(\omega)}(x) & \cdots & \phi_{K}^{(\omega)}(x)  \end{array} \right)_{\omega\times (K\times \omega)}.
\end{equation*}


The dynamic functional principal component analysis is an effective way of extracting a large amount of long-run covariance. It summarizes the main features of an infinite-dimensional object by its finite elements by minimizing the mean integrated squared error of the reconstruction error function over the whole functional data set. The dynamic functional principal component analysis has been applied in modeling human age-specific mortality \citep{shang2019} and foreign exchange rates \cite{SF21}. This work is the first that extends the technique to model multivariate functional time series to the best of our knowledge. The dynamic multivariate functional time series method could better capture the temporal dependency of age-specific mortality rates compared to the commonly used Lee-Carter model or static functional time series method \citep[see, e.g.,][]{HU07}.

\subsection{Long-run covariance estimation} \label{3.2}

The long-run covariance function introduced in~\eqref{eq_1} in practice can be estimated from a finite sample $\bm{f}(x) = \left\{\bm{f}_1(x), \dots, \bm{f}_n(x) \right\}$. A natural estimator of $C(x,s)$ in a bi-infinite sum is given by
\begin{equation}
  \widehat{C}_{v,q}(x,s) = \sum_{\ell = -\infty}^{\infty} W_q\left(\frac{\ell}{v}\right) \widehat{\gamma}_{\ell}(x,s), \label{eq_3}
\end{equation}
where $v$ is the bandwidth parameter, and 
\begin{equation*}
 \widehat{\gamma}_{\ell} (x,s) =
\begin{cases}
\frac{1}{n} \sum_{t=1}^{n-\ell} \left[\bm{f}_t(x) - \widehat{\bm{\mu}}(x) \right] \left[\bm{f}_{t+\ell}(s) - \widehat{\bm{\mu}}(s) \right], & \ell \geq 0; \\
\frac{1}{n} \sum_{t=1-\ell}^{n} \left[\bm{f}_t(x) - \widehat{\bm{\mu}}(x) \right] \left[\bm{f}_{t+\ell}(s) - \widehat{\bm{\mu}}(s) \right], & \ell < 0,
\end{cases}
\end{equation*}
with $\widehat{\bm{\mu}}(x) = n^{-1} \sum_{t=1}^{n} \bm{f}_t(x) $. The weight function $W_q$ in~\eqref{eq_3} is a continuous and symmetric function with bounded support of order $q$ defined on $[-m,m]$ for some $m>0$ satisfying
\begin{equation*}
  W_q(0) = 1, W_q(u) \leq 1, W_q(u) = W_q(-u), W_q(u) = 0 \text{ if } |u| > m, 
\end{equation*}
and there exists an $\omega$ such that
\begin{equation*}
  0 < \omega = \operatornamewithlimits{lim}_{u \rightarrow 0} |u|^{-q}(1-W_q(u)) < \infty. 
\end{equation*}
As with the kernel sandwich estimator in~\eqref{eq_3}, the crucial part is the estimation of bandwidth parameter $v$. We estimated $v$ through a data-driven approach using the plug-in algorithm of \citet[][Section 2]{RS17}. The plug-in algorithm selects the optimal bandwidth $v$ that minimizes the asymptotic mean-squared normed error between the estimated and actual long-run variance functions.

\subsection{Selecting the number of retained functional principal components} \label{sec:3.3}

There are several methods for selecting the number of functional principal components $K$: 
\begin{inparaenum}
\item[(1)] the fraction of the variance explained by the first few functional principal components \citep{Chiou12};
\item[(2)] pseudo-versions of the Akaike information criterion (AIC) and Bayesian information criterion \citep{YMW05};
\item[(3)] predictive cross-validation leaving one or more curves out \citep{RS91}; and
\item[(4)] bootstrap methods \citep{HV06}.
\end{inparaenum}
Here, the number of components is determined as the minimum that reaches 90\% of total variance explained by the leading components (the first method mentioned), such that
\begin{equation*}
K = \argmin_{K: K\geq 1}\left\{\sum^K_{k=1}\widehat{\lambda}_k\Bigg/\sum^{\infty}_{k=1}\widehat{\lambda}_k\mathds{1}_{\left\{\widehat{\lambda}_k>0\right\}}\geq 0.9\right\},
\end{equation*}
where $\mathds{1}_{\{\widehat{\lambda}_k>0\}}$ is to exclude possible zero eigenvalues, and $\mathds{1}\{\cdot\}$ represents the binary indicator function. Although the total variance explained level could be arbitrary, 0.9 is a common threshold when using the total variance explained in principal component analysis to select the number of components to retain \citep[see also][]{KRX21}.

\subsection{Forecasting multivariate functional time series}\label{sec:3.4}

In the dynamic multivariate functional time series method, we first stack all series into a vector of functions and obtain a set of dynamic functional principal components with associated scores according to the method described in Section~\ref{sec:3.1}. Conditioning on the estimated mean function and estimated dynamic functional principal components, these forecast scores are multiplied by the estimated principal components to form forecast curves. A univariate time series forecasting method, such as the ARIMA model, is considered to forecast dynamic functional principal component scores. Through differencing, the ARIMA model can handle non-stationarity present in the data. For details of forecasting dynamic principal component scores, consult \cite{HS09}, \cite{shang2019} and \cite{SF21}. 

Denote the $h$-step-ahead prediction of dynamic principal component scores $\left\{\beta_1^{(l)}, \cdots, \beta_K^{(l)}\right\}$ by $\left\{\widehat{\beta}_{n+h,1}^{(l)},\cdots,\widehat{\beta}_{n+h,K}^{(l)}\right\}$. We construct the $h$-step-ahead point forecast of $f_{n+h}(x)$ as
\begin{equation*}
  \widehat{f}_{n+h}^{(l)}(x) = \text{E}\left[f_{n+h}^{(l)}(x)|\bm{f}^{(l)}(x), \bm{B}\right] = \widehat{\mu}^{(l)}(x) + \sum_{k=1}^{K} \widehat{\beta}_{n+h,k}^{(l)} \widehat{\phi}^{(l)}(x),
\end{equation*}
where $\bm{B} = \left\{\widehat{\phi}_1^{(l)}(x), \dots,\widehat{\phi}_K^{(l)}(x) \right\}$ are the estimated functional principal component functions.

\subsection{Construction of pointwise prediction intervals} \label{sec:3.5}

As a supplement to point forecast evaluation, prediction intervals are constructed and evaluated to gauge the uncertainty associated with point forecasts. Depending on statistical theory and data on error distributions, prediction intervals explicitly estimate the probability that future realizations lie within a given range. In implementing the proposed mortality forecasting method, the errors accumulated in the estimated principal component scores and model residuals in~\eqref{eq_2} are the primary sources of uncertainty. As emphasized by~\cite{Chatfield93}, it is crucial to provide interval forecasts as well as point forecasts to be able to
\begin{inparaenum}[(1)]
\item assess future uncertainty level;
\item enable different strategies to be planned for the range of possible outcomes;
\item compare forecasts from different methods; and
\item explore different scenarios based on various underlying assumptions.
\end{inparaenum}

To construct uniform and pointwise prediction intervals, we adopt the method of~\cite{ANH15}. The method can be summarized in the following steps:
\begin{enumerate}[1)]
\item Using all observed data, compute the $K$-variate score vectors $\Big(\bm{\beta}_1^{(l)}, \cdots, \bm{\beta}_K^{(l)}\Big)$ and the sample functional principal components $\Big[\widehat{\phi}_1^{(l)}(x), \cdots, \widehat{\phi}_K^{(l)}(x)\Big]$. Then, we can construct in-sample forecasts
\begin{equation*}
\widehat{f}_{\xi+h}^{(l)}(x) = \widehat{\beta}_{\xi+h,1}^{(l)}\widehat{\phi}_1^{(l)}(x) +\cdots + \widehat{\beta}_{\xi+h,K}^{(l)}\widehat{\phi}_K^{(l)}(x),
\end{equation*}
where $\Big(\widehat{\beta}_{\xi+h,1}^{(l)},\cdots,\widehat{\beta}_{\xi+h,K}^{(l)} \Big)$ are the elements of the $h$-step-ahead prediction obtained from $\Big(\bm{\beta}_1^{(l)}, \cdots, \bm{\beta}_K^{(l)}\Big)$ by a univariate time series forecasting method, for $\xi \in \{K, K+1, \cdots,n-h\}$.
\item With the in-sample point forecasts, we calculate the in-sample point forecast errors
\begin{equation*}
\widehat{\epsilon}_{\zeta}^{(l)}(x) = f_{\xi+h}^{(l)}(x) - \widehat{f}_{\xi+h}^{(l)}(x),
\end{equation*}
where $\zeta \in \{1,2,\cdots, M \} $ and $M = n-h-K+1$.
\item Based on these in-sample forecast errors, we can sample with replacement to obtain a series of bootstrapped forecast errors, from which we obtain lower and upper bounds, denoted by $\gamma^{(l)}_{\text{lb}}(x)$ and $\gamma^{(l)}_{\text{ub}}(x)$, respectively. We then seek a tuning parameter $\pi_{\alpha}$ such that $\alpha \times 100\%$ of the residuals satisfy
\begin{equation*}
\pi_{\alpha} \times \gamma^{(l)}_{\text{lb}}(x_i) \leq \widehat{\epsilon}_{\zeta}^{(l)}(x_i) \leq \pi_{\alpha} \times \gamma^{(l)}_{\text{ub}}(x_i), \qquad i = 1,\dots,p.
\end{equation*}
Then, the $h$-step-ahead pointwise prediction intervals are given as 
\begin{equation*}
\pi_{\alpha} \times \gamma^{(l)}_{\text{lb}}(x_i) \leq f_{n+h}^{(l)}(x_i) - \widehat{f}_{n+h}^{(l)}(x_i) \leq \pi_{\alpha} \times \gamma^{(l)}_{\text{ub}}(x_i),
\end{equation*}
where $i$ symbolizes the discretized data points. 
\end{enumerate}

Note that~\cite{ANH15} calculate the standard deviation of $\Big[\widehat{\epsilon}_{1}^{(l)}(x), \cdots, \widehat{\epsilon}_{M}^{(l)}(x)\Big]$, which leads to a parametric approach to constructing prediction intervals. Instead, we consider the nonparametric bootstrap approach of~\cite{Shang17}, which allows us to reconcile bootstrapped forecasts among different functional time series in a hierarchy.

\subsection{Reconciliation of grouped forecasts} \label{sec:3.6}

The disaggregated data can first be organized into groups for forecasting sub-national mortality rates before applying the dynamic multivariate functional time series method. Hence, an appropriate grouping structure is crucial for accurate forecasts. A natural strategy is arranging the sub-national functional time series into geographical structures, as in Figure~\ref{fig:4}. State-of-the-art forecast reconciliation methods are applied to ensure that the point and interval forecasts at the various disaggregation levels add up to the forecasts of the national mortality.

For all aforementioned hierarchies, we denote a particular disaggregated series using the notation $\text{G} \ast \text{S}$, meaning the geographical area G and the sex S. For instance, $\text{R}_1 \ast \text{F}$ denotes females in Region 1, $\text{P}_1 \ast \text{T}$ denotes females and males in Prefecture 1 and $\text{Japan} \ast \text{M}$ denotes males in Japan. Let $E_{\text{G}\ast \text{S}, t}(x)$ denote the exposure to risk for series $\text{G}\ast \text{S}$ in year $t$ and age $x$, and let $D_{\text{G}\ast \text{S}, t}(x)$ be the number of deaths for series $\text{G}\ast \text{S}$ in year $t$ and age $x$. Then, age-specific mortality rate is given by $R_{\text{G}\ast\text{S}, t}(x) = D_{\text{G}\ast \text{S}, t}(x)/E_{\text{G}\ast \text{S}, t}(x)$. Dropping the age argument $(x)$ allows us to express the national and sub-national mortality series in a matrix multiplication as
\arraycolsep=0.04cm
\[
\underbrace{ \left[
\begin{array}{l}
R_{\text{Japan}\ast \text{T},t} \\
R_{\textcolor{red}{\text{Japan}\ast \text{F},t}} \\
R_{\textcolor{red}{\text{Japan}\ast \text{M},t}} \\
R_{\textcolor{a0}{\text{R1}\ast \text{T},t}} \\
R_{\textcolor{a0}{\text{R2}\ast \text{T},t}} \\
\vdots \\
R_{\textcolor{a0}{\text{R8}\ast \text{T},t}} \\
R_{\textcolor{blue-violet}{\text{R1}\ast \text{F},t}} \\
R_{\textcolor{blue-violet}{\text{R2}\ast \text{F},t}} \\
\vdots \\
R_{\textcolor{blue-violet}{\text{R8}\ast \text{F},t}} \\
R_{\textcolor{burntorange}{\text{R1}\ast \text{M},t}} \\
R_{\textcolor{burntorange}{\text{R2}\ast \text{M},t}} \\
\vdots \\
R_{\textcolor{burntorange}{\text{R8}\ast \text{M},t}} \\
R_{\textcolor{blue}{\text{P1}\ast \text{T},t}} \\
R_{\textcolor{blue}{\text{P2}\ast \text{T},t}} \\
\vdots \\
R_{\textcolor{blue}{\text{P47}\ast \text{T},t}} \\
R_{\textcolor{purple}{\text{P1}\ast \text{F},t}} \\
R_{\textcolor{blush}{\text{P1}\ast \text{M},t}} \\
R_{\textcolor{purple}{\text{P2}\ast \text{F},t}} \\
R_{\textcolor{blush}{\text{P2}\ast \text{M},t}} \\
\vdots \\
R_{\textcolor{purple}{\text{P47}\ast \text{F},t}} \\
R_{\textcolor{blush}{\text{P47}\ast \text{M},t}} \\ \end{array}
\right]}_{\bm{R}_t} =
\underbrace{\left[
\begin{array}{ccccccccccc}
\frac{E_{\text{P1}\ast \text{F},t}}{E_{\text{Japan}\ast \text{T},t}} & \frac{E_{\text{P1}\ast \text{M},t}}{E_{\text{Japan}\ast \text{T},t}} & \frac{E_{\text{P2}\ast \text{F},t}}{E_{\text{Japan}\ast \text{T},t}} & \frac{E_{\text{P2}\ast \text{M},t}}{E_{\text{Japan}\ast \text{T},t}}  & \frac{E_{\text{P3}\ast \text{F},t}}{E_{\text{Japan}\ast \text{T},t}} & \frac{E_{\text{P3}\ast \text{M},t}}{E_{\text{Japan}\ast \text{T},t}} & \cdots & \frac{E_{\text{P47}\ast \text{F},t}}{E_{\text{Japan}\ast \text{T},t}} & \frac{E_{\text{P47}\ast \text{M},t}}{E_{\text{Japan}\ast \text{T},t}} \\
\textcolor{red}{\frac{E_{\text{P1}\ast \text{F},t}}{E_{\text{Japan}\ast \text{F},t}}} & \textcolor{red}{0} & \textcolor{red}{\frac{E_{\text{P2}\ast \text{F},t}}{E_{\text{Japan}\ast \text{F},t}}} & \textcolor{red}{0} & \textcolor{red}{\frac{E_{\text{P3}\ast \text{F},t}}{E_{\text{Japan}\ast \text{F},t}}} & \textcolor{red}{0} & \cdots & \textcolor{red}{\frac{E_{\text{P47}\ast \text{F},t}}{E_{\text{Japan}\ast \text{F},t}}} & \textcolor{red}{0} \\
\textcolor{red}{0} & \textcolor{red}{\frac{E_{\text{P1}\ast \text{M},t}}{E_{\text{Japan}\ast \text{M},t}}}  & \textcolor{red}{0} & \textcolor{red}{\frac{E_{\text{P2}\ast \text{M},t}}{E_{\text{Japan}\ast \text{M},t}}} & \textcolor{red}{0} & \textcolor{red}{\frac{E_{\text{P3}\ast \text{M},t}}{E_{\text{Japan}\ast \text{M},t}}} & \cdots & \textcolor{red}{0} & \textcolor{red}{\frac{E_{\text{P47}\ast \text{M},t}}{E_{\text{Japan}\ast \text{M},t}}} \\
\textcolor{a0}{\frac{E_{\text{P1}\ast \text{F},t}}{E_{\text{R1}\ast \text{T},t}}} & \textcolor{a0}{\frac{E_{\text{P1}\ast \text{M},t}}{E_{\text{R1}\ast \text{T},t}}} & \textcolor{a0}{0} & \textcolor{a0}{0} & \textcolor{a0}{0} & \textcolor{a0}{0} & \cdots  & \textcolor{a0}{0} & \textcolor{a0}{0} \\
\textcolor{a0}{0} & \textcolor{a0}{0} & \textcolor{a0}{\frac{E_{\text{P2}\ast \text{F},t}}{E_{\text{R2} \ast \text{T},t}}} & \textcolor{a0}{\frac{E_{\text{P2}\ast \text{M},t}}{E_{\text{R2}\ast \text{T},t}}} & \textcolor{a0}{\frac{E_{\text{P3}\ast \text{F},t}}{E_{\text{R2}\ast \text{T},t}}} & \textcolor{a0}{\frac{E_{\text{P3}\ast \text{M},t}}{E_{\text{R2}\ast \text{T},t}}} & \cdots & \textcolor{a0}{0} & \textcolor{a0}{0} \\
\vdots & \vdots & \vdots & \vdots & \vdots & \vdots & \cdots & \vdots & \vdots \\
\textcolor{a0}{0} & \textcolor{a0}{0} & \textcolor{a0}{0} & \textcolor{a0}{0} & \textcolor{a0}{0} & \textcolor{a0}{0} & \cdots & \textcolor{a0}{\frac{E_{\text{P47}\ast \text{F},t}}{E_{\text{R8}\ast \text{T},t}}} & \textcolor{a0}{\frac{E_{\text{P47}\ast \text{M},t}}{E_{\text{R8}\ast \text{T},t}}} \\
\textcolor{blue-violet}{\frac{E_{\text{P1}\ast \text{F},t}}{E_{\text{R1}\ast \text{F},t}}} & \textcolor{blue-violet}{0} & \textcolor{blue-violet}{0} & \textcolor{blue-violet}{0} & \textcolor{blue-violet}{0} & \textcolor{blue-violet}{0} &  \cdots & \textcolor{blue-violet}{0} & \textcolor{blue-violet}{0} \\
\textcolor{blue-violet}{0} & \textcolor{blue-violet}{0} & \textcolor{blue-violet}{\frac{E_{\text{P2}\ast \text{F},t}}{E_{\text{R2}\ast \text{F},t}}} & \textcolor{blue-violet}{0} & \textcolor{blue-violet}{\frac{E_{\text{P3}\ast \text{F},t}}{E_{\text{R2}\ast \text{F},t}}} & \textcolor{blue-violet}{0} & \cdots & \textcolor{blue-violet}{0} & \textcolor{blue-violet}{0}  \\
\vdots & \vdots & \vdots & \vdots & \vdots & \vdots & \cdots & \vdots & \vdots \\
\textcolor{blue-violet}{0} & \textcolor{blue-violet}{0}  & \textcolor{blue-violet}{0}  & \textcolor{blue-violet}{0}  & \textcolor{blue-violet}{0}  & \textcolor{blue-violet}{0}  & \cdots & \textcolor{blue-violet}{\frac{E_{\text{P47}\ast \text{F},t}}{E_{\text{R8}\ast \text{F},t}}} & \textcolor{blue-violet}{0}\\
\textcolor{burntorange}{0} & \textcolor{burntorange}{\frac{E_{\text{P1}\ast \text{M},t}}{E_{\text{R1}\ast \text{M},t}}} & \textcolor{burntorange}{0} &\textcolor{burntorange}{0} & \textcolor{burntorange}{0} & \textcolor{burntorange}{0} & \cdots & \textcolor{burntorange}{0} & \textcolor{burntorange}{0} \\
\textcolor{burntorange}{0} & \textcolor{burntorange}{0} & \textcolor{burntorange}{0} & \textcolor{burntorange}{\frac{E_{\text{P2}\ast \text{M},t}}{E_{\text{R2}\ast \text{M},t}}} & \textcolor{burntorange}{0} & \textcolor{burntorange}{\frac{E_{\text{P3}\ast \text{M},t}}{E_{\text{R2}\ast \text{M},t}}} & \cdots & \textcolor{burntorange}{0} & \textcolor{burntorange}{0} \\
\vdots & \vdots & \vdots & \vdots & \vdots & \vdots & \cdots & \vdots & \vdots \\
\textcolor{burntorange}{0} & \textcolor{burntorange}{0} & \textcolor{burntorange}{0} & \textcolor{burntorange}{0} & \textcolor{burntorange}{0} & \textcolor{burntorange}{0} & \cdots & \textcolor{burntorange}{0} & \textcolor{burntorange}{\frac{E_{\text{P47}\ast \text{M},t}}{E_{\text{R8}\ast \text{M},t}}} \\
\textcolor{blue}{\frac{E_{\text{P1}\ast \text{F},t}}{E_{\text{P1}\ast \text{T},t}}} & \textcolor{blue}{\frac{E_{\text{P1}\ast \text{M},t}}{E_{\text{P1}\ast \text{T},t}}} & \textcolor{blue}{0} & \textcolor{blue}{0} & \textcolor{blue}{0} & \textcolor{blue}{0} & \cdots & \textcolor{blue}{0} & \textcolor{blue}{0} \\
\textcolor{blue}{0} & \textcolor{blue}{0}  &  \textcolor{blue}{\frac{E_{\text{P2}\ast \text{F},t}}{E_{\text{P2}\ast \text{T},t}}} & \textcolor{blue}{\frac{E_{\text{P2}\ast \text{M},t}}{E_{\text{P2}\ast \text{T},t}}} & \textcolor{blue}{0} & \textcolor{blue}{0}  & \cdots & \textcolor{blue}{0} & \textcolor{blue}{0} \\
\vdots & \vdots & \vdots & \vdots & \vdots & \vdots & \cdots & \vdots & \vdots \\
\textcolor{blue}{0} & \textcolor{blue}{0} & \textcolor{blue}{0} & \textcolor{blue}{0} & \textcolor{blue}{0} & \textcolor{blue}{0} & \cdots & \textcolor{blue}{\frac{E_{\text{P47}\ast \text{F},t}}{E_{\text{P47}\ast \text{T},t}}} & \textcolor{blue}{\frac{E_{\text{P47}\ast \text{M},t}}{E_{\text{P47}\ast \text{T},t}}} \\
\textcolor{purple}{1} & \textcolor{purple}{0} & \textcolor{purple}{0} & \textcolor{purple}{0} & \textcolor{purple}{0} & \textcolor{purple}{0} & \cdots & \textcolor{purple}{0} & \textcolor{purple}{0} \\
\textcolor{blush}{0} & \textcolor{blush}{1} & \textcolor{blush}{0} & \textcolor{blush}{0} & \textcolor{blush}{0} & \textcolor{blush}{0} & \cdots & \textcolor{blush}{0} & \textcolor{blush}{0} \\
\textcolor{purple}{0} & \textcolor{purple}{0} & \textcolor{purple}{1} & \textcolor{purple}{0} & \textcolor{purple}{0} & \textcolor{purple}{0} & \cdots & \textcolor{purple}{0} & \textcolor{purple}{0} \\
\textcolor{blush}{0} & \textcolor{blush}{0} & \textcolor{blush}{0} & \textcolor{blush}{1} & \textcolor{blush}{0} & \textcolor{blush}{0} & \cdots & \textcolor{blush}{0} & \textcolor{blush}{0} \\
\vdots & \vdots & \vdots & \vdots & \vdots & \vdots & \cdots & \vdots & \vdots  \\
\textcolor{purple}{0} & \textcolor{purple}{0} & \textcolor{purple}{0} & \textcolor{purple}{0} & \textcolor{purple}{0} & \textcolor{purple}{0} & \cdots  & \textcolor{purple}{1} & \textcolor{purple}{0}\\
\textcolor{blush}{0} & \textcolor{blush}{0} & \textcolor{blush}{0} & \textcolor{blush}{0} & \textcolor{blush}{0} & \textcolor{blush}{0} & \cdots & \textcolor{blush}{0} & \textcolor{blush}{1} \\
\end{array}
\right]}_{\bm{S}_t}
\underbrace{\left[
\begin{array}{l}
R_{\text{P1}\ast \text{F},t} \\
R_{\text{P1}\ast \text{M},t} \\
R_{\text{P2}\ast \text{F},t} \\
R_{\text{P2}\ast \text{M},t} \\
\vdots \\
R_{\text{P47}\ast \text{F},t} \\
R_{\text{P47}\ast \text{M},t} \\
\end{array}
\right]}_{\bm{b}_t}
\]
\hspace{-.05in} or $\bm{R}_t = \bm{S}_t\bm{b}_t$, where $\bm{R}_t$ is a vector containing all series at all levels of disaggregation, $\bm{b}_t$ is a vector of the most disaggregated series, and $\bm{S}_t$ shows how the two are related. The group structures illustrated in Figure~\ref{fig:4} satisfy this relationship. In the two group structures, the main difference is about how the $(R_{R1\times F, t}$, $R_{R1\times M, t}$,$\dots$, $R_{R8\times F, t}$, $R_{R8\times M, t})$ and $(R_{P1\times F, t}$, $R_{P1\times M, t}$, $\dots$, $R_{P47\times F, t}$, $R_{P47\times M, t}$) are estimated. In Figure~\ref{fig:4}\subref{fig:4b}, we estimate the female and male series together in a geographical location. In Figure~\ref{fig:4}\subref{fig:4c}, we estimate multiple series at various geographical locations for females and males, respectively.

We present a brief review of three methods for forecast reconciliation based on this summing matrix. The three forecast reconciliation methods are bottom-up, optimal combination method with the ordinary least squares (OLS) estimator, and optimal combination method with generalized least squares (GLS) estimator. The bottom-up method has the agreeable feature that it is intuitive and straightforward and always results in forecasts that satisfy the same group structure as the original data \citep[e.g., see][]{DM92, ZT00}. Instead of considering only time series in the bottom-level, \cite{HAA+11} proposed a method in which base forecasts for all aggregated and disaggregated series are computed successively. Then, the resulting forecasts are combined through linear regression. With the OLS estimator, the reconciled forecasts are close to the base forecasts and aggregate consistently within the group. \cite{WAH19} proposed an improved approach of finding coherent forecasts for the grouped functional time series. By minimizing the sum of variances of reconciliation errors, essentially minimizing trace (MinT) of the variance matrix, this approach produces coherent forecasts across the entire collection of time series. Since the variance-covariance matrix of out-of-sample reconciliation errors is often unknown and not identifiable in practice, the MinT method attempts to approximate weights with the in-sample base forecast errors using the GLS estimator.

\section{Data analysis results} \label{sec:result}

We apply the dynamic univariate and multivariate functional time series forecasting methods to Japanese age-specific mortality rates to obtain base forecasts for the group structures in Figure~\ref{fig:4}. Then, we conduct reconciliation via the bottom-up and the optimal combination methods. To assess model and parameter stabilities over time, we consider an expanding window analysis of considered time series models \citep[see][Chapter 9 for details]{ZW06}. We used the observed mortality curves in 1975--2001 Japanese data to produce one- to 15-step-ahead point and interval forecasts. Through an expanding window approach, we re-estimate the parameters of considered models using the first 28 years of observations from 1975 to 2002 and estimate one- to 14-step-ahead forecasts using re-estimated models. The process is iterated with the sample size increased by one year each time until reaching the end of the data period in 2016. This process produces 15 one-step-ahead forecasts, 14 two-step-ahead forecasts, and so on, up to one 15-step-ahead forecast. We evaluate the point and interval forecast accuracies and report our study results in this section.

\subsection{Point forecast evaluation and comparison} \label{sec:5.1}

To evaluate the point forecast accuracy, we compute the root mean squared forecast error (RMSFE), which measures how close the forecasts are compared to the actual values. For each series, the RMSFE is given by
\begin{align*}
\text{RMSFE}(h) &= \sqrt{\frac{1}{101\times (16-h)}\sum^{15}_{\varsigma = h}\sum^{101}_{i=1}\left[f_{n+\varsigma}(x_i) - \widehat{f}_{n+\varsigma}(x_i)\right]^2},
\end{align*}
where $f_{n+\varsigma}(x_i)$ represents the actual holdout sample for the $i$\textsuperscript{th} age and $\varsigma$\textsuperscript{th} curve of the forecasting period, and $\widehat{f}_{n+\varsigma}(x_i)$ is the corresponding point forecasts. We obtain measures of point forecast accuracies for all all national and sub-national mortality series, averaging over 15 forecast horizons, as
\begin{equation*}
\text{Mean (RMSFE)} = \frac{1}{15}\sum^{15}_{h=1}\text{RMSFE}(h).
\end{equation*}

\begin{figure}[!htb]
\centering
\subfloat{\includegraphics[width = 6.5in]{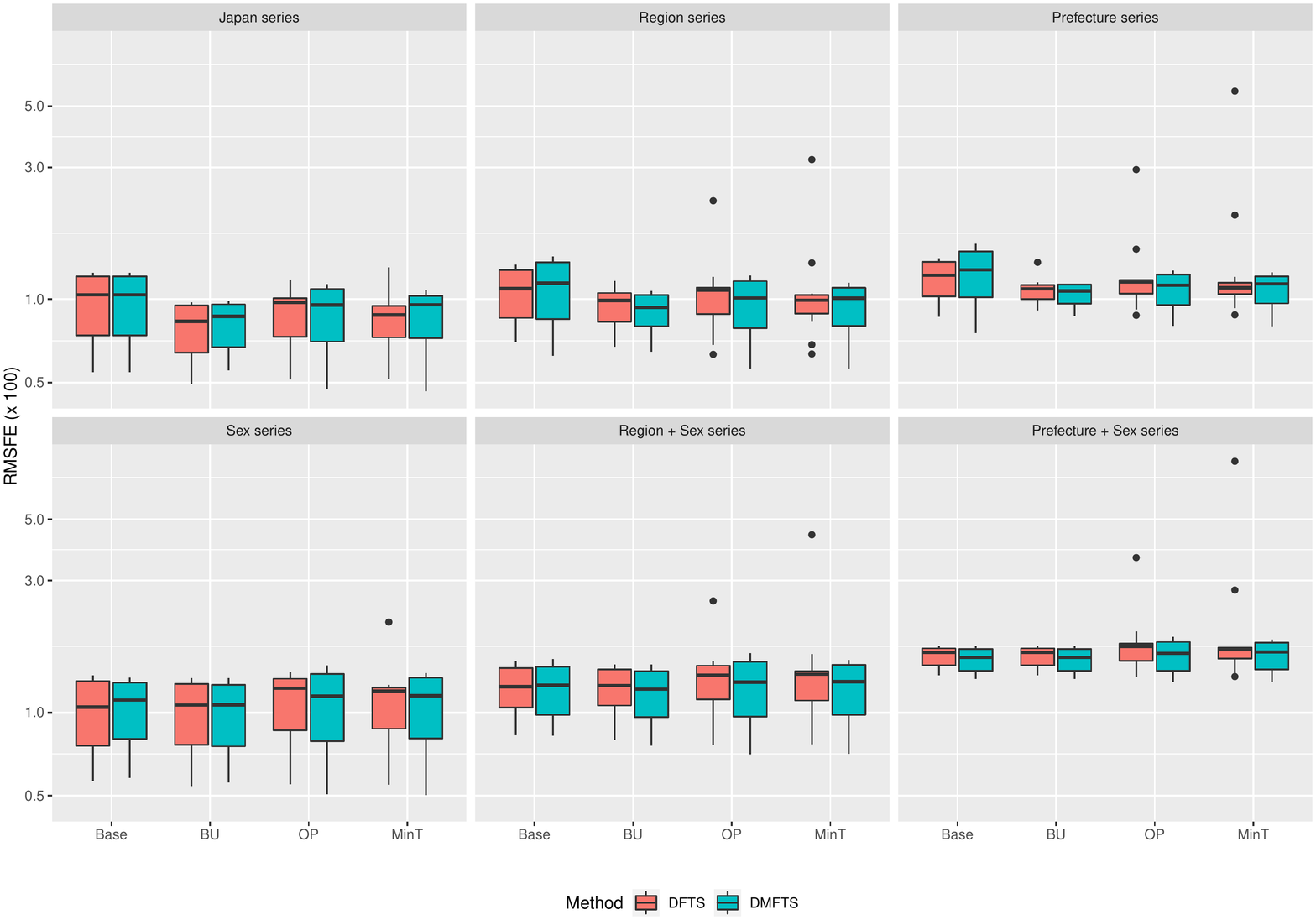}} 
\caption{Point forecasting performance of the independent (Base) and reconciled forecasts (BU, OP, and MinT) for univariate and multivariate functional time series methods. DFTS denotes the dynamic univariate functional time series method, while DMFTS denotes the dynamic multivariate functional time series method.}
\label{fig:5}
\end{figure}

Averaging all series at each disaggregation level of Hierarchy~1, we show RMSFEs ($\times 100$) for the forecasts before and after reconciliations in Figure~\ref{fig:5}. In the figure, we denote the independent forecasts as Base. We denote the reconciled forecasts obtained by the bottom-up method and the optimal combination method as BU and OP, respectively \citep[see, e.g.,][for more details]{HAA+11,Shang16}. This figure compares forecasts obtained by the dynamic multivariate functional time series method with forecasts produced by its univariate counterpart method. The dynamic multivariate functional time series method is expected to give more accurate point forecasts at disaggregation levels where multiple mortality series have highly consistent patterns. The correlations among multiple series can be captured and used in producing forecasts. This advantage of our method is shown in ``Region $+$ Sex series'' and ``Prefecture $+$ Sex series'' panels of Figure~\ref{fig:5}. However, due to its construction, the dynamic multivariate functional time series method implicitly assumes the same number of retained functional principal components for all series, consequently resulting in diminishing forecast accuracy for a particular series. This disadvantage of our method is revealed when only the $\text{Japan} \ast \text{F}$ and $\text{Japan} \ast \text{M}$ series are considered, as shown in the ``Sex series'' panel of Figure~\ref{fig:5}. 
 
At the national level, where only one mortality time series exists, the dynamic univariate functional time series method, a special case of dynamic multivariate functional time series method, is implemented based on an estimated long-run covariance function. Hence, both methods have the same forecasts for the total Japanese series. 

The advantage of dynamic multivariate functional time series method is that it could extract common features among strongly related sub-populations. Another prominent character of the dynamic multivariate functional time series method shown in Figure~\ref{fig:5} is the method's robustness to outliers in the training sample. At least three prefectures in Japan that were severely affected by earthquakes and tsunamis have inflated mortality curves. One reason for the dynamic multivariate functional time series forecasts being immune to outliers is that the computation of long-run covariance function over a period mitigates the influence of abrupt and temporal movements of observations. The autocovariance function, a part of the long-run covariance function, can smooth out the observational noise. Another reason for the robust forecast is that collectively, the consideration of related sub-populations enables the information pooling and thus improves the extraction of human mortality features. By contrast, many large values exist in the dynamic univariate functional time series forecasts even after reconciliation.

Figure~\ref{fig:5} also shows that pairing all three reconciliation methods with either dynamic multivariate or univariate functional time series method improves the point forecast accuracy. Among the reconciliation methods considered, the bottom-up method yields the most accurate overall point forecasts when used with dynamic multivariate functional time series, owing to the high signal-to-noise ratio of the prefecture-level sex-specific mortality series. Our results confirm \citeauthor{SH16}'s \citeyearpar{SH16} early finding regarding the superiority of the bottom-up method for the Japanese sub-national age-specific mortality rates.

\begin{figure}[!htb]
\centering
\subfloat{\includegraphics[width = 6.5in]{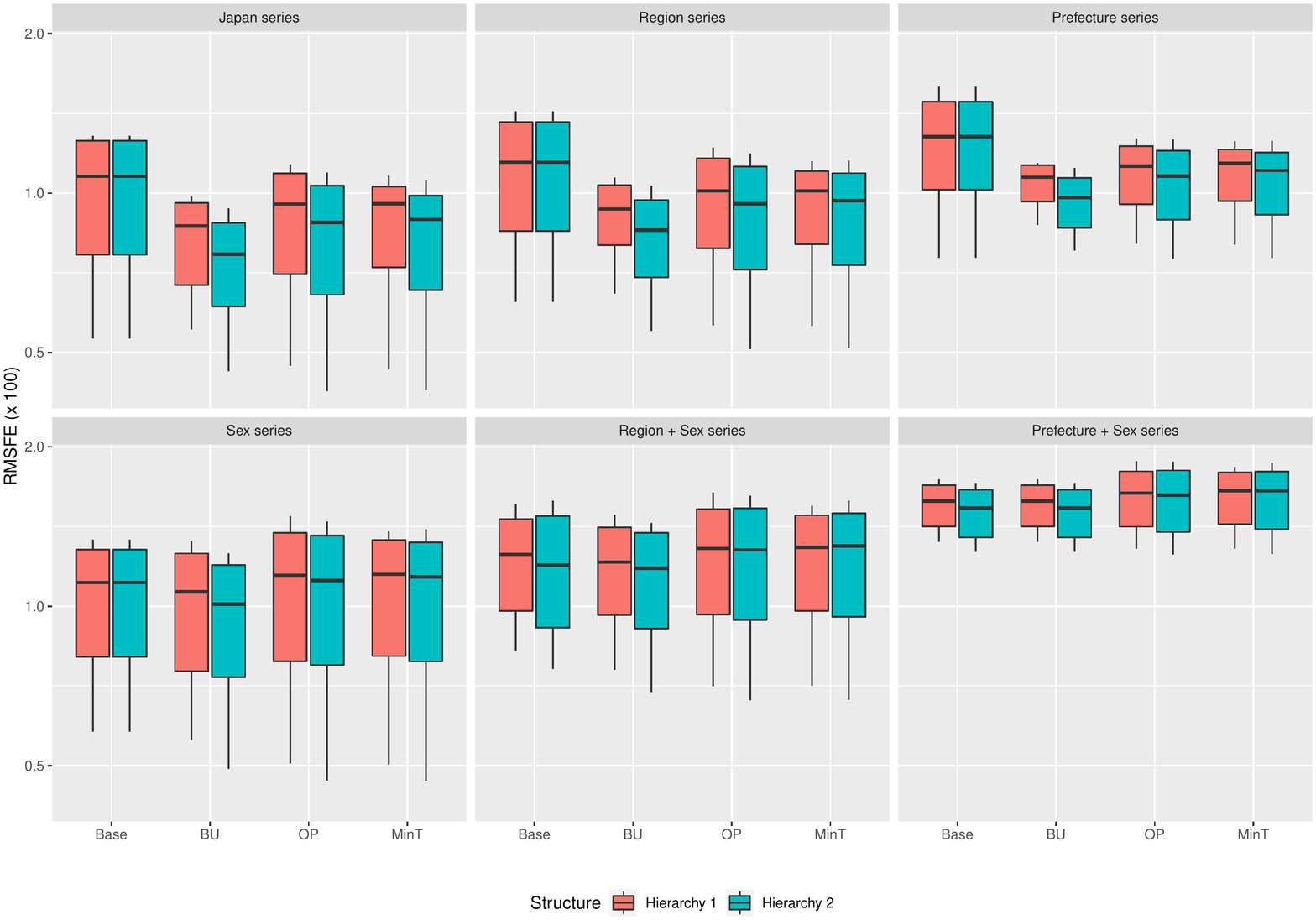}} 
\caption{RMSFEs ($\times 100$) for the base and reconciled multivariate functional time series forecasts obtained from different group structures.}
\label{fig:6}
\end{figure}

Applying the dynamic multivariate functional time series method to various disaggregation structures leads to different RMSFEs ($\times 100$), as shown in Figure~\ref{fig:6}. The geographical hierarchies described in Figure~\ref{fig:4} are blended into Hierarchy~1 and Hierarchy~2 in the reconciliation of the total series at the country, region, and prefecture levels. Hierarchy~2 produces the most accurate point forecasts, given the outliers in the data. Collectively modeling the curves that represent all females (males) in the prefectures can better ease the influence of the few outlying curves than jointly modeling the series for females and males in each prefecture. This finding is because natural disasters are likely to affect females and males in an area equally. Comparing RMSFEs across all sub-national levels, we recommend grouping age-specific mortality rates of the same sex for each region and prefecture, applying the dynamic multivariate functional time series method to obtain base forecasts, and then reconciling the forecasts using the bottom-up method. 

\subsection{Interval forecast evaluation and comparison} \label{sec:5.2}

In addition to point forecasts, we also evaluate the pointwise interval forecast accuracy using the interval score of \cite{GR07} \citep[see also][]{Gneiting14}. For each year in the forecasting period, the $h$-step-ahead prediction intervals are calculated at the $100(1-\alpha)\%$ nominal coverage probability. We consider the common case of a symmetric $100(1-\alpha)\%$ prediction interval whose lower and upper bounds are predictive quantiles at $\alpha/2$ and $1-\alpha/2$, denoted by $\widehat{f}_{n+\varsigma}^{lb}(x_i)$ and $\widehat{f}_{n+\varsigma}^{ub}(x_i)$, respectively. As defined by \cite{GR07}, a scoring rule for the pointwise interval forecast at time point $x_i$ is
\begin{align*} 
P_{\alpha} \left[ \widehat{f}_{n+\varsigma}^{lb}(x_i), \widehat{f}_{n+\varsigma}^{ub}(x_i); f_{n+\varsigma}(x_i)  \right] =&  \left[ \widehat{f}_{n+\varsigma}^{ub}(x_i) -  \widehat{f}_{n+\varsigma}^{lb}(x_i)  \right]  \\
&+ \frac{2}{\alpha}\left[ \widehat{f}_{n+\varsigma}^{lb}(x_i) - f_{n+\varsigma}(x_i)  \right] \mathbbm{1}\left\lbrace f_{n+\varsigma}(x_i) < \widehat{f}_{n+\varsigma}^{lb}(x_i)  \right\rbrace\\
& + \frac{2}{\alpha}\left[  f_{n+\varsigma}(x_i) -  \widehat{f}_{n+\varsigma}^{ub}(x_i)  \right] \mathbbm{1} \left\lbrace f_{n+\varsigma}(x_i) >  \widehat{f}_{n+\varsigma}^{ub}(x_i)   \right\rbrace,
\end{align*}
where $\alpha$ denotes the level of significance; typically $\alpha = 0.2$. It can be easily seen that the optimal interval score is achieved when $f_{n+\varsigma}(x_i)$ lies between $\widehat{f}_{n+\varsigma}^{lb}(x_i)$ and $\widehat{f}_{n+\varsigma}^{ub}(x_i)$, with the distance between the upper bound and the lower bound being minimal.

The scoring rule of \cite{GR07} is considered in this study since it combines the coverage rate and halfwidth of the prediction interval. We aim to construct prediction intervals where the empirical coverage rate is close to the nominal one, with a minimum width of the prediction interval. When the empirical coverage rate is less than the nominal one, the constructed prediction interval is narrower than the oracle. This indicates that we are under-estimating the forecast uncertainty. Conversely, if the empirical coverage rate is more than the nominal one, our prediction interval is wider than the oracle. This indicates we are over-estimating the forecast uncertainty, and our prediction intervals are not that informative.

Averaging over different points in a curve and different years in the forecasting period, the mean interval score is defined as
\begin{align*}
\overline{P}_{\alpha}(h) = \frac{1}{101\times (16-h)}\sum^{15}_{\varsigma = h}\sum^{101}_{i=1}P_{\alpha} \left[ \widehat{f}_{n+\varsigma}^{lb}(x_i), \widehat{f}_{n+\varsigma}^{ub}(x_i); f_{n+\varsigma}(x_i)  \right], 
\end{align*}
where $P_{\alpha}\left[ \widehat{f}_{n+\varsigma}^{lb}(x_i), \widehat{f}_{n+\varsigma}^{ub}(x_i); f_{n+\varsigma}(x_i)  \right]$ denotes the interval score at the $\varsigma^{th}$ curve of the forecasting period. For 15 different forecast horizons, we use a mean statistic given by
\begin{equation*}
\text{Mean}(\overline{P}_{\alpha}) = \frac{1}{15} \sum_{h=1}^{15}\overline{P}_{\alpha}(h)
\end{equation*}
to evaluate the pointwise interval forecast accuracy.

\begin{figure}[!htb]
	\centering
	\subfloat{\includegraphics[width = 6.6in]{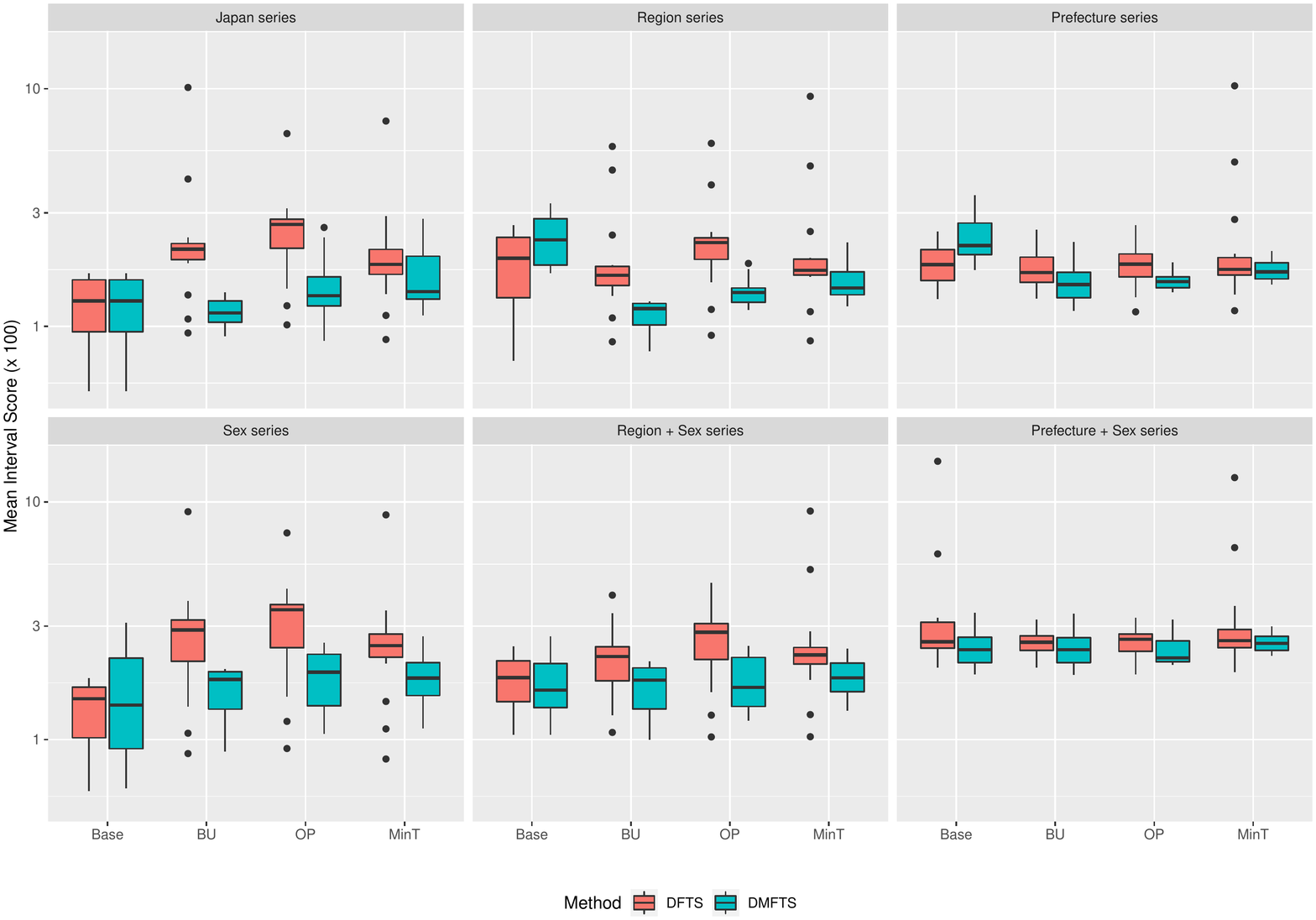}} 
	\caption{Interval forecasting performance of the base and reconciled forecasts for the univariate and multivariate functional time series methods.}
	\label{fig:7}
\end{figure}

Figure~\ref{fig:7} summarizes the mean interval scores of the univariate and multivariate functional time series applied to the Japanese age-specific mortality rates under Hierarchy~1. Considering multiple mortality curves allows for borrowing information of related populations but at the expense of sacrificing specific information for each sub-population to a certain extent. Hence, the multivariate forecasting method may not always outperform the univariate forecasting method because of the relatively large heterogeneity of the considered mortality series. In our study, pointwise interval forecasts obtained by the dynamic multivariate functional time series method are generally more accurate than those produced by the dynamic univariate functional time series method; the superior forecast accuracy of the dynamic multivariate functional time series method becomes more evident if used in combination with the bottom-up method.

The second and third panels in Figure~\ref{fig:7} depict interval forecast accuracies for the total series (i.e., the aggregate mortality of females and males) at region and prefecture levels, respectively. The DMFTS method collectively models all the considered total series to estimate the long-run covariance of the entire population of Japan before producing forecasts. Significant heterogeneity exists in the total series at the Prefecture or Region levels since earthquakes and tsunamis influenced mortality at several prefectures during the considered period. Hence, natural disaster-related deaths as outliers tend to inflate the estimation variance of every series used in the DMFTS modeling process. In contrast, the base method considers one total series, either for a region or a prefecture, at a time. Only those prefectures or regions with outlying death counts would see additional estimation variance. Moreover, the dynamic modeling method may not accurately recover the long-run covariance of mortality series of all Japanese residents based on a finite sample. Due to these reasons, the DMFTS method sometimes performs inferiorly to the independent method. Applying any of the considered reconciliation methods improves interval forecast accuracies of the total series at region and prefecture levels, as shown in Figure~\ref{fig:7}. The forecast accuracy improvement is because Japan's population hierarchy helps stabilize the abnormal forecasts inflated by excess uncertainty in estimation.

\begin{figure}[!htb]
	\centering
	\subfloat{\includegraphics[width = 6.48in]{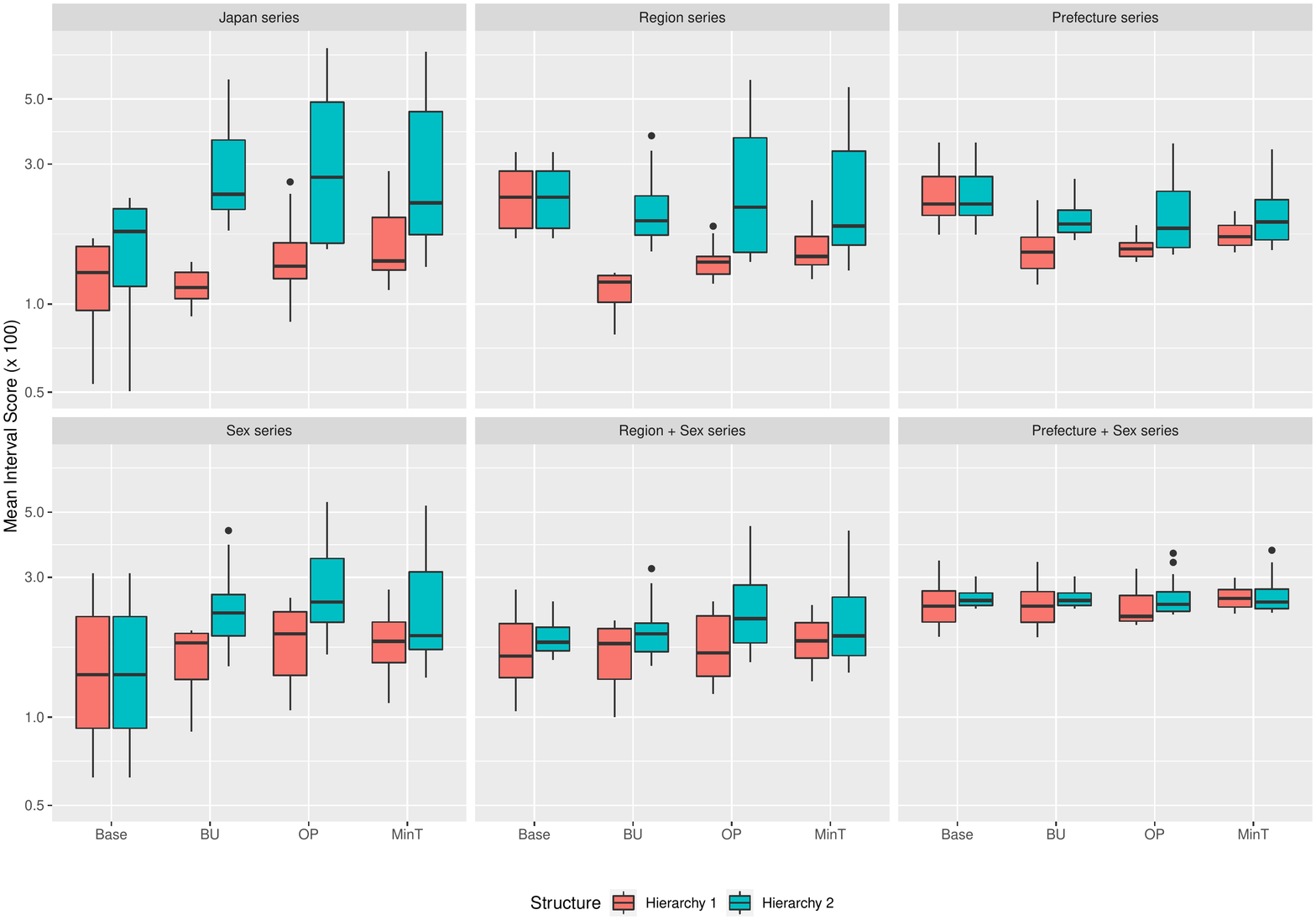}} 
	\caption{Mean interval scores for the base and reconciled multivariate functional time series forecasts obtained from different group structures.}
	\label{fig:8}
\end{figure}

Figure~\ref{fig:8} compares the mean interval scores for the base and reconciled multivariate functional time series forecasts obtained using two disaggregation hierarchies. The result of Hierarchy~1 dominating Hierarchy~2, shown in the figure, can be explained as follows. By applying the dynamic multivariate functional time series method under the Hierarchy~2 structure, it focuses on common features of female (male) mortality, leading to inaccurate pointwise prediction intervals for outlying series due to losing specific information for the sub-population. Given that interval scores will be inflated by every holdout value outside the calculated prediction interval, outlier curves cause a greater penalty to Hierarchy~2 than to Hierarchy~1. Hence, jointly modeling the mortality series for females (males) of each prefecture and region is better for improving pointwise interval forecast accuracy.

\section{Conclusion} \label{sec:conclusion}

When applied to the functional time series formed by disaggregated series, such as sub-national age-specific mortality rates, the dynamic multivariate functional time series method can be combined with reconciliation methods to obtain improved point and interval forecasts. 

The dynamic multivariate functional time series method is applied to data on Japanese age-specific mortality rates from 1975 to 2016. The dynamic multivariate functional time series method produces the most accurate point and interval forecasts of Japanese mortality rates combined with the bottom-up reconciliation method. Different group structures involving geographical factors and sex are considered and compared in forecasting and reconciliation. For point forecasting, we recommend that, first, the national mortality data are disaggregated by sex, and further, each series is split according to a geographical area. Because of the bias-variance trade-off, reversing the disaggregation order, that is, first, according to the geographical area, and then, by sex, yields more accurate pointwise interval forecasts, evaluated by the mean interval scores. 

There are some ways in which this paper can be further extended. Here, we briefly mention three:
\begin{enumerate}
\item[1)] We have considered two factors, namely sex and geographical locations, to disaggregate age-specific mortality rates. It is possible to enrich the information contained in the group structure by differentiating these rates according to other factors, such as the cause of death \citep{ML97, GS15}, social and economic aspects \citep{BBA02, SAS+13} and native-born or immigrant status \citep{SGS01}. If appropriate data are available, we may attempt to extend the proposed grouped multivariate functional time series forecasting method to cause-specific mortality rates and occupation-specific mortality rates.
\item[2)] We expect the mortality functions at the prefecture-level to be homogeneous so that the empirical long-run covariance function can be accurately estimated. When there are major natural disasters such as earthquakes and tsunamis in some prefectures, the observed mortality functions may have different shapes. Such functions of strange shapes can be viewed as outliers. Functional depth-based outlier detection methods of \cite{HS10}, and the recently developed novel multivariate functional outlier detection methods of \cite{DG18} and~\cite{DG20} can be adopted to check if a particular curve is an outlying observation. Among other methods, the robust functional principal component analysis method and the robust regularized singular value decomposition method of \cite{SFY2019} can be applied to mitigate influences of functional outliers in modeling and forecasting age-specific morality rates \citep[see also][]{ZSH13,BJB+11}.
\item[3)] We evaluate and compare the pointwise interval forecast accuracy. To our best knowledge, uniform prediction intervals are still rarely used in the context of functional time series models, owing to the difficulty in computing an appropriate semi-metric distance. A future extension can consider the uniform prediction intervals of grouped functional time series and study the interval forecast accuracy in function space.
\end{enumerate}

\section*{Supplementary Material}
For reproducibility, all R codes are collated in a Github repository at \url{https://github.com/hanshang/GMFTS}. This repository contains the following files:
\begin{itemize}
 \item \textbf{README}: Describe the functionality of each R file.
 \item \textbf{R files with indices 1 to 29}: Functions used to compute numerical results presented in the paper.
 \item \textbf{Example}: Contains a quick example of forecasting prefecture-level age-specific mortality rates using the DMFTS method.
\end{itemize}



\bibliographystyle{agsm}
\bibliography{GFTS}

\end{document}